\documentclass[sigconf,balance=false]{acmart}

\usepackage{textcomp}
\usepackage{stfloats}
\usepackage{graphicx}
\usepackage{url}
\usepackage{verbatim}
\usepackage{microtype}
\usepackage{subcaption}
\usepackage{booktabs}
\usepackage{balance}

\usepackage{amsmath}

\usepackage{amssymb}
\usepackage{mathtools}
\usepackage{amsthm}

\usepackage{float} 
\usepackage{multirow,enumitem}

\theoremstyle{plain}

\theoremstyle{definition}

\theoremstyle{remark}

\DeclareCaptionType{algorithm}[Algorithm][List of Algorithms]

\usepackage{color, colortbl}
\definecolor{Gray}{rgb}{0.9, 0.9, 0.9}

\usepackage{xcolor}

\setlist{nosep,leftmargin=*}
\renewcommand{\arraystretch}{0.96}

\renewcommand\footnotetextcopyrightpermission[1]{} 
\AtBeginDocument{%
  }

\setcopyright{acmlicensed}
\copyrightyear{2024}
\acmYear{2024}
\acmDOI{XXXXXXX.XXXXXXX}

\acmConference[Conference acronym 'XX]{Make sure to enter the correct
  conference title from your rights confirmation emai}{June 03--05,
  2018}{Woodstock, NY}
\begin{document}

\title{Agents as Knowledge Integrator and Utilizer in Multimodal Recommendation}

\author{Jinfeng Xu}
\affiliation{%
  \institution{The University of Hong Kong}
  \city{Hong Kong SAR}
  \country{China}
}
\email{jinfeng@connect.hku.hk}

\author{Zheyu Chen}
\affiliation{%
  {\institution{Beijing Institute of Technology}}
  \city{Beijing}
  \country{China}}
\email{zheyu.chen@bit.edu.cn}

\author{Shuo Yang}
\affiliation{%
  \institution{The University of Hong Kong}
  \city{Hong Kong SAR}
  \country{China}
}
\email{shuoyang.ee@gmail.com}

\author{Jinze Li}
\affiliation{%
  \institution{The University of Hong Kong}
  \city{Hong Kong SAR}
  \country{China}
}
\email{lijinze-hku@connect.hku.hk}


\author{Puzhen Wu}
\affiliation{%
  \institution{The University of Hong Kong}
  \city{Hong Kong SAR}
  \country{China}
}
\email{puzhenwu8@connect.hku.hk}

\author{Zewei Liu}
\affiliation{%
  \institution{The University of Hong Kong}
  \city{Hong Kong SAR}
  \country{China}
}
\email{zewliu@hku.hk}

\author{Zheng Lin}
\affiliation{%
  \institution{The University of Hong Kong}
  \city{Hong Kong SAR}
  \country{China}
}
\email{linzheng@eee.hku.hk}

\author{Jianheng Tang}
\affiliation{%
  \institution{Peking University}
  \city{Beijing}
  \country{China}
}
\email{tangentheng@gmail.com}

\author{Jing Yang}
\affiliation{%
  \institution{Universiti Malaya}
  \city{Kuala Lumpur}
  \country{Malaysia}
}
\email{s2147529@siswa.um.edu.my}

\author{Wei Wang}
\affiliation{%
  \institution{Macao Polytechnic University}
  \city{Macao SAR}
  \country{China}
}
\email{weiwang@mpu.edu.mo}

\author{Xiping Hu}
\affiliation{%
  \institution{Beijing Institute of Technology}
  \city{Beijing}
  \country{China}
}
\email{huxp@bit.edu.cn}

\author{Edith Ngai}
\authornote{Corresponding Author.}
\affiliation{%
  \institution{The University of Hong Kong}
  \city{Hong Kong SAR}
  \country{China}
}
\email{chngai@eee.hku.hk}

\renewcommand{\shortauthors}{Xu et al.}

\begin{abstract}
Online platforms increasingly rely on multimodal recommender systems to rank products, media, and other Web content. Existing methods usually inject visual and textual features into item representations or build homogeneous graphs from modality-level similarity, but the resulting signals can remain misaligned with the recommendation objective. We study this semantic gap from a knowledge-integration perspective: multimodal content should be interpreted together with user behavior before it is used to construct recommendation graphs or adjust rankings.

We propose AgentMMRec, an agent-based multimodal recommendation framework with two coordinated roles. The Integrator Agent infers behavior- and multimodal-aware user preferences and item properties from training interactions and item content, then stores them in a reusable knowledge memory. The Utilizer Agent consumes this memory to refine modality-specific item-item graphs, construct behavior-aware homogeneous graphs, and rerank candidate lists under a frozen evaluation-time memory. This design differs from direct LLM feature augmentation and pure LLM reranking because the generated knowledge is first converted into graph structure and model representations before recommendation. Experiments on three Amazon multimodal recommendation datasets show that AgentMMRec consistently improves Recall and NDCG over recent multimodal baselines, remains effective under sparsity and item cold-start settings, and can transfer its constructed knowledge to existing backbones.
\end{abstract}



\maketitle

\section{Introduction}
\label{sec:introduction}
The growth of online product catalogs has made multimodal recommendation a central Web application: items are described by text, images, brands, categories, and user interaction histories, while recommenders must convert these heterogeneous signals into ranked lists. Recent multimodal recommenders enrich item representations with visual and textual features or build homogeneous graphs from modality-level similarity \citep{xu2025best,zhou2023comprehensive,xu2025enhancing,zhou2023tale,xu2025mdvt,zhang2022latent}. These designs improve the representation space, but they do not fully resolve the semantic gap between multimodal content and the recommendation objective \citep{xu2025survey,liu2024multimodal}. Item images and descriptions may encode color, style, material, or marketing language that is visually or textually salient but weakly related to why users interact with the item. A graph constructed from such features can therefore connect items that look similar while missing behaviorally relevant relations.

LLM-based recommendation provides a new way to interpret item content together with behavioral context \citep{wei2024llmrec,ren2024representation,fioretti2025powerful,xu2025enhancing,bao2023tallrec,xu2026dggvae}. Existing studies mainly use LLMs for data augmentation \citep{xi2023kar,wei2024llmrec}, as fine-tuned recommendation backbones \citep{bao2023tallrec,zhang2025collm}, or as list rerankers \citep{hou2023large,hou2024large,zhang2025ur4rec}. These paradigms are useful but incomplete for multimodal recommendation. Augmentation still produces content-side features, fine-tuning is costly and data hungry, graph construction is often decoupled from reusable user/item memory, and pure reranking only adjusts a late-stage candidate list. What remains missing is a mechanism that converts LLM-derived semantic knowledge into reusable recommendation structure while preserving a standard recommender training and evaluation protocol.

We propose AgentMMRec, a two-agent framework that treats LLM-derived knowledge as an intermediate memory rather than as a direct replacement for the recommender. The Integrator Agent extracts behavior- and multimodal-aware user preferences and item properties from training interactions and item content, then stores them in a key-value knowledge memory. The Utilizer Agent uses this memory to refine modality-specific item-item graphs, construct behavior-aware homogeneous graphs, and rerank recommendation candidates. The key distinction is not the use of an LLM alone, but the role assigned to its output: the same frozen memory is converted into graph structure, representation input, and controlled candidate reranking. This design separates AgentMMRec from LLM-generated profiling, graph-only enhancement, and LLM-only reranking baselines.

Experiments on real-world datasets show that AgentMMRec improves over recent multimodal recommendation baselines across all metrics. The gains persist in sparsity and item cold-start settings, and the compatibility study shows that the constructed graphs and knowledge memory can improve several existing multimodal recommenders. We further analyze memory updating, LLM backbone choice, self-supervised objectives, template robustness, and cost. The main contributions are:
\begin{itemize}[leftmargin=*]
    \item We identify the semantic gap between multimodal item content and recommendation objectives, and we distinguish this gap from the feature augmentation and pure reranking views adopted by existing LLM-based recommenders.
    \item We propose AgentMMRec, which uses an Integrator Agent to construct behavior-aware knowledge memory and a Utilizer Agent to transform that memory into refined homogeneous graphs, representation enhancement, and candidate reranking.
    \item We evaluate AgentMMRec on multiple multimodal recommendation datasets and analyze transferability, sparsity, cold-start behavior, memory updating, LLM backbone choice, SSL compatibility, template robustness, and computational cost.
\end{itemize}

\section{Related Work}
\label{sec:related work}
\subsection{Multimodal Recommendation}
Multimodal recommendation uses item text, images, and interactions to alleviate sparsity and improve ranking. VBPR \citep{he2016vbpr} incorporated visual content into matrix factorization \citep{rendle2009bpr}, and later models combined visual and textual modalities to enrich item representations \citep{chen2019personalized,liu2019user,yu2023multi,chen2025don,lin2025contrastive,chen2025causality}. Graph-based methods further exploit user-item connectivity and modality similarity. MMGCN \citep{wei2019mmgcn} extracts modality-specific signals through graph convolution, DualGNN \citep{wang2021dualgnn} and LATTICE \citep{zhang2021mining} introduce user-user or item-item graphs, and FREEDOM \citep{zhou2023tale} improves representation stability by freezing item semantic graphs and denoising the user-item graph. Recent methods also use self-supervised learning and cross-modal alignment. MMSSL \citep{wei2023multi}, BM3 \citep{zhou2023bootstrap}, and MENTOR \citep{xu2025mentor} align multimodal inputs with collaborative signals, while LGMRec \citep{guo2024lgmrec} and COHESION \citep{xu2025cohesion} model higher-order or composite graph structure.

These methods show that modality information is useful, but they still rely heavily on feature-level similarity. Recent surveys identify the semantic gap between multimodal data and recommendation tasks as a persistent challenge \citep{xu2025survey,liu2024multimodal}. AgentMMRec addresses this gap by interpreting multimodal content through behavioral context before using it to build homogeneous graphs and enhanced representations.

\subsection{LLM-based Recommendation}
LLMs have been used to enhance recommendation through content understanding, user profiling, item attribute extraction, and list reranking \citep{xi2023kar,wei2024llmrec,tian2023graph,bao2023tallrec,hou2023large,lee2024star,zhang2025llm,ren2025easyrec,zhang2025collm,zhang2025ur4rec}. TALLRec \citep{bao2023tallrec} uses instruction tuning with LLaMA \citep{touvron2023llama}, LEARN \citep{zhang2025llm} converts item attributes into prompts and uses LLM hidden states as item embeddings, and LLMRank \citep{hou2023large} reranks retrieved items conditioned on interaction history. KAR \citep{xi2023kar} and LLMRec \citep{wei2024llmrec} use LLM-derived user preferences and item attributes to address sparse feedback and weak side information. UR4Rec \citep{zhang2025ur4rec} further retrieves user preferences and item knowledge for candidate-aware reranking.

Recent multimodal studies are closer to our setting because they use LLMs or MLLMs to reason over item modalities and behavior. DOGE \citep{meng2025doge} builds an LLM-enhanced hyper-knowledge graph for multimodal recommendation, while MLLMRec \citep{dang2026mllmrec} introduces MLLM-driven preference reasoning and graph refinement. AgentMMRec differs in the way the generated knowledge is organized and reused. Instead of treating LLM output only as augmented text, item embeddings, a graph-construction signal, or a final reranking score, AgentMMRec stores preference and property knowledge in a memory that is used for three linked operations: behavior-aware homogeneous graph construction, representation enhancement, and frozen-memory candidate reranking. This design lets the same knowledge benefit standard multimodal recommenders while keeping the recommender backbone and the LLM components modular.

\begin{figure*}[t]
    \centering
    \includegraphics[width=1\linewidth]{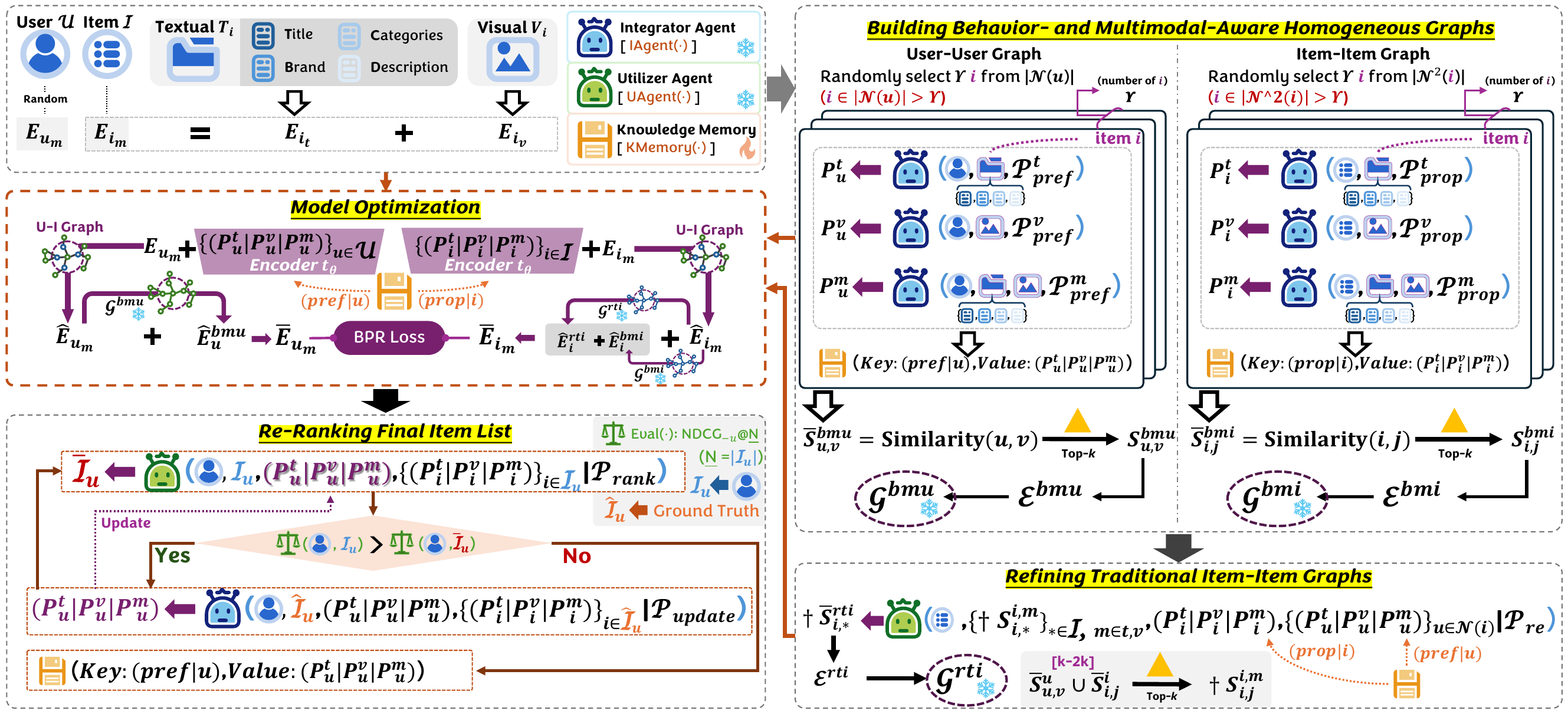}
    \vskip -0.1in
    \caption{Overview of AgentMMRec. The Integrator Agent builds behavior-aware knowledge memory from training interactions and item content, while the Utilizer Agent converts this memory into graph refinement, representation enhancement, and candidate reranking.}
    \label{fig:overview}
    \vskip -0.1in
\end{figure*}

\section{Methodology}
\label{sec:methodology}
As illustrated in Figure~\ref{fig:overview} and Algorithm~\ref{algorithm:process}, AgentMMRec contains an Integrator Agent ($\operatorname{IAgent}(\cdot)$), a Utilizer Agent ($\operatorname{UAgent}(\cdot)$), and a knowledge memory. The Integrator Agent interprets item multimodal content under user-behavior context and writes behavior-aware user preferences and item properties into memory. The Utilizer Agent consumes the stored knowledge to refine modality-specific item-item graphs, build behavior- and multimodal-aware homogeneous graphs, enhance recommender representations, and rerank candidate lists. In this framework, an agent is an LLM/VLM module with a defined input state, prompt-guided action, memory read/write scope, and stopping rule. The two-agent design separates knowledge construction from knowledge use: extracted knowledge is reusable across training epochs and can be transferred to other multimodal recommendation backbones.

\subsection{Problem Definition}
Formally, let $\mathcal{U} = \{u_1,\ldots,u_{|\mathcal{U}|}\}$ and $\mathcal{I} = \{i_1,\ldots,i_{|\mathcal{I}|}\}$ be the user and item sets. Each item $i$ contains textual fields, including title $T_i^{title}$, brand $T_i^{brand}$, categories $T_i^{categories}$, and description $T_i^{description}$, together with an image $V_i$. Following representative multimodal recommendation models \citep{chen2025don,zhou2023tale,guo2024lgmrec,xu2025survey}, we use the MMRec encoders \citep{zhou2023mmrecsm} for fair comparison: a text encoder $t_\theta(\cdot)$ and a visual encoder $v_\theta(\cdot)$. The textual representation is $\mathbf{e}_{i_t}=t_\theta(T_i)$, where $T_i=(T_i^{\text{title}}|T_i^{\text{brand}}|T_i^{\text{categories}}|T_i^{\text{description}})$ and $|$ denotes concatenation. The visual representation is $\mathbf{e}_{i_v}=v_\theta(V_i)$. Item representations for modality $m\in\{t,v\}$ are denoted by $\mathbf{E}_{i_m}\in\mathbb{R}^{d_m\times|\mathcal{I}|}$, and user representations are randomly initialized as $\mathbf{E}_{u_m}\in\mathbb{R}^{d_m\times|\mathcal{U}|}$. The interaction matrix is $\mathcal{R}\in\{0,1\}^{|\mathcal{U}|\times|\mathcal{I}|}$, where $\mathcal{R}_{u,i}=1$ indicates an observed interaction. It induces the bipartite graph $\mathcal{G}=(\mathcal{U},\mathcal{I},\mathcal{E})$ with bidirectional edges for observed pairs. Because the studied multimodal recommendation setting has no temporal behavior sequence, the central challenge is to infer user preferences and item properties from sparse interactions and multimodal item content.

\subsection{Building Behavior- and Multimodal-Aware Homogeneous Graphs}
The Integrator Agent extracts user preferences and item properties by combining training interactions with multimodal data, then stores the extracted knowledge in memory. The stored knowledge is used to build a behavior- and multimodal-aware user-user graph and a behavior- and multimodal-aware item-item graph.

\subsubsection{Behavior-Aware User-User Graph}
\label{sec:bmuug}
Users usually lack direct multimodal profiles, so prior multimodal recommenders either omit user-user homogeneous graphs or build them only from interaction patterns \citep{wang2021dualgnn,xu2025cohesion,zhou2023comprehensive,xu2025survey}. The Integrator Agent instead generates behavior- and multimodal-aware preferences $P_u$ for each user $u \in \mathcal{U}$ by interpreting the multimodal content of training items that $u$ interacted with. Formally:
\begin{equation}
    P_u^t \leftarrow \operatorname{IAgent}(u, \{T_i^{\text{t}}, T_i^{\text{b}}, T_i^{\text{c}}, T_i^{\text{d}}\}_{i \in \mathcal{N}(u)}|\mathcal{P}_{pref}^t),
\end{equation}
\begin{equation}
    P_u^v \leftarrow \operatorname{IAgent}(u, \{V_i\}_{i \in \mathcal{N}(u)}|\mathcal{P}_{pref}^v),
\end{equation}
\begin{equation}
    P_u^m \leftarrow \operatorname{IAgent}(u, \{T_i^{\text{t}}, T_i^{\text{b}}, T_i^{\text{c}}, T_i^{\text{d}}, V_i\}_{i \in \mathcal{N}(u)}|\mathcal{P}_{pref}^m),
\end{equation}
where $\mathcal{N}(u)$ denotes user $u$'s observed training interactions. For users with more than $\Upsilon$ interacted items, we sample $\Upsilon$ items to control LLM context length and inference cost. Users may express modality-specific preferences, such as brand or color, and cross-modal preferences, such as a brand-color combination. We therefore extract textual, visual, and cross-modal preferences with three prompt templates; the full templates are provided with the anonymous code package. The extracted preferences are stored in the knowledge memory as key-value entries:
\begin{equation}
    \operatorname{KMemory}(\text{Key}:(pref|u), \text{Value}: (P_u^t| P_u^v| P_u^m)).
\end{equation}
Subsequently, we use pre-trained encoder $t_\theta(\cdot)$ to compute the representation of user preferences and construct a top-$k$ behavior- and multimodal-aware user-user graph $\mathcal{G}^{bmu} = (\mathcal{U},\mathcal{E}^{bmu})$ based on cosine similarity. Formally:
\begin{equation}
    \mathcal{S}_{u,v}^{bmu}=\left\{\begin{array}{ll}
1 & \text { if } \mathcal{\bar{S}}_{u,v}^{bmu} \in \text {top-} k\left(\mathcal{\bar{S}}_{u,*}^{bmu}\right) \\
0 & \text { otherwise }
\end{array},\right. 
\end{equation}
\begin{equation}
\mathcal{\bar{S}}_{u,v}^{bmu} = \frac{t_\theta(P_u^t| P_u^v| P_u^m)^{T}t_\theta(P_v^t| P_v^v| P_v^m)}{\|t_\theta(P_u^t| P_u^v| P_u^m)\|\|t_\theta(P_v^t| P_v^v| P_v^m)\|}.
\end{equation}

Then, we build row-wise top-$k$ directed edges $(u,v) \in \mathcal{E}^{bmu}$, where $\mathcal{S}_{u,v}^{bmu} = 1$. During representation enhancement, $\mathcal{D}^{bmu}$ denotes the row-degree matrix of this adjacency.

\subsubsection{Behavior-Aware Item-Item Graph}
\label{sec:bmiig}
Many existing multimodal recommendation models construct item-item graphs based on multimodal data. Our AgentMMRec also incorporates a refined traditional item-item graph (refer to Section~\ref{sec:rtiig}). However, directly constructing an item-item graph using item features focuses only on the multimodal data itself, without considering the specific requirements of the recommendation task.

Users who interact with the same items often share preference signals, and their co-purchased items can reveal behavior-aware item properties. For item $i$, the Integrator Agent interprets the multimodal data of other training items purchased by users who interacted with $i$ and generates behavior- and multimodal-aware properties $P_i$. Formally:
\begin{equation}
    P_i^t \leftarrow \operatorname{IAgent}(i, \{T_i^{\text{t}}, T_i^{\text{b}}, T_i^{\text{c}}, T_i^{\text{d}}\}_{i \in \mathcal{N}^2(i)}|\mathcal{P}_{prop}^t),
\end{equation}
\begin{equation}
    P_i^v \leftarrow \operatorname{IAgent}(i, \{V_i\}_{i \in \mathcal{N}^2(i)}|\mathcal{P}_{prop}^v),
\end{equation}
\begin{equation}
    P_i^m \leftarrow \operatorname{IAgent}(i, \{T_j^{\text{t}}, T_j^{\text{b}}, T_j^{\text{c}}, T_j^{\text{d}}, V_j\}_{j \in \mathcal{N}^2(i)}|\mathcal{P}_{prop}^m),
\end{equation}
where $\mathcal{N}^2(i)$ denotes other training items purchased by users who interacted with item $i$. If $|\mathcal{N}^2(i)|>\Upsilon$, we sample $\Upsilon$ items. As with user preferences, the Integrator Agent extracts textual, visual, and cross-modal properties, and the complete prompt templates are included with the code (The code will be made publicly available upon acceptance of the paper). The item properties are stored as:
\begin{equation}
    \operatorname{KMemory}(\text{Key}:(prop|i), \text{Value}: (P_i^t| P_i^v| P_i^m)).
\end{equation}
Subsequently, we use pre-trained encoder $t_\theta(\cdot)$ to compute the representation of item properties and construct a top-$k$ behavior- and multimodal-aware item-item graph $\mathcal{G}^{bmi} = (\mathcal{I},\mathcal{E}^{bmi})$ based on cosine similarity. Formally, this process can be expressed as:
\begin{equation}
    \mathcal{S}_{i,j}^{bmi}=\left\{\begin{array}{ll}
1 & \text { if } \mathcal{\bar{S}}_{i,j}^{bmi} \in \text {top-} k\left(\mathcal{\bar{S}}_{i,*}^{bmi}\right) \\
0 & \text { otherwise }
\end{array},\right. 
\end{equation}
\begin{equation} \mathcal{\bar{S}}_{i,j}^{bmi} = \frac{t_\theta(P_i^t| P_i^v| P_i^m)^{T}t_\theta(P_j^t| P_j^v| P_j^m)}{\|t_\theta(P_i^t| P_i^v| P_i^m)\|\|t_\theta(P_j^t| P_j^v| P_j^m)\|}.
\end{equation}

Then, we build row-wise top-$k$ directed edges $(i,j) \in \mathcal{E}^{bmi}$, where $\mathcal{S}_{i,j}^{bmi} = 1$. During representation enhancement, $\mathcal{D}^{bmi}$ denotes the row-degree matrix of this adjacency.

\textbf{Discussion. } The behavior- and multimodal-aware homogeneous graphs are constructed before recommender training, so graph construction does not add per-epoch graph-building cost. The LLM call volume is reported separately in Section~\ref{sec:efficiency}. The threshold $\Upsilon$ controls both context length and inference cost, and its sensitivity is analyzed in Section~\ref{sec:hyperparameter}.

\subsection{Refining Traditional Item-Item Graphs}
\label{sec:rtiig}
Traditional multimodal recommendation models \citep{zhang2022latent,xu2025cohesion,zhou2023tale} construct modality-specific item-item graphs based on item representations to enhance modality representations. However, this process exacerbates the isolation between modalities \citep{xu2025best} and lacks consideration of user preferences. Utilizer Agent leverages the behavior- and multimodal-aware preferences and properties stored in the constructed knowledge memory to refine and merge modality-specific item-item graphs into a unified item-item graph. Specifically, original modality-specific item-item graphs are constructed as:
\begin{equation}
    \dagger\mathcal{S}_{i,j}^{i,m}=\left\{\begin{array}{ll}
1 & \text { if } \dagger\mathcal{\bar{S}}_{i,j}^{i,m} \in \text {top-} k\left(\dagger\mathcal{\bar{S}}_{i,*}^{i,m}\right) \\
0 & \text { otherwise }
\end{array},\right. 
\end{equation}
\begin{equation}
\dagger\mathcal{\bar{S}}_{i,j}^{i,m} = \frac{(\mathbf{e}_{i_m})^{T}\mathbf{e}_{j_m}}{\|\mathbf{e}_{i_m}\|\|\mathbf{e}_{j_m}\|},
\end{equation}
where $m\in\{t,v\}$. For each modality, we construct a top-$k$ modality-specific item-item graph. For item $i$, the Utilizer Agent then combines its text and visual neighbors, item properties, and the preferences of users who purchased it. Guided by $\mathcal{P}_{re}$, the Utilizer Agent reselects the top-$k$ items and constructs a unified item-item graph $\mathcal{G}^{rti}=(\mathcal{I},\mathcal{E}^{rti})$:
\begin{equation}
\begin{aligned}
    \{\dagger \mathcal{S}_{i,*}^{rti}\}_{*\in \mathcal{I}}
    &\leftarrow \operatorname{UAgent}\bigl(\\
    &i,\{\dagger\mathcal{S}_{i,*}^{i,m}\}_{\substack{*\in \mathcal{I}\\m \in \{t,v\}}},
    (P_i^t|P_i^v|P_i^m),\\
    &\{(P_u^t|P_u^v|P_u^m)\}_{u \in \mathcal{N}(i)}
    \mid \mathcal{P}_{re}\bigr),
\end{aligned}
\end{equation}
where $\mathcal{N}(i)$ denotes the purchased user set for item $i$ and the number of selected neighbors is constrained by $\sum_{j\in \mathcal{I}}\dagger\mathcal{S}_{i,j}^{rti}=k$. We also adopt $\Upsilon$ to constrain the size of the purchased user set $\mathcal{N}(i)$. Then, we build row-wise directed edges $(i,j) \in \mathcal{E}^{rti}$, where $\dagger\mathcal{S}_{i,j}^{rti} = 1$. The implementation asks the Utilizer Agent to return a ranked item-id list; invalid ids are discarded before the adjacency is saved.

\textbf{Discussion. } The refined item-item graph is also constructed before recommender training. This keeps the training loop close to standard graph-based recommenders, while the offline LLM cost is measured separately.

\subsection{Representation Enhancement and Model Optimization}
\label{sec:rep_enhancement}
We enhance user and item representations by leveraging encoded behavior- and multimodal-aware preferences and properties. Following the paradigm adopted by most previous studies \citep{xu2025survey,zhou2023comprehensive}, we apply LightGCN \citep{he2020lightgcn} to propagate messages and perform readout over the user-item interaction graph $\mathcal{G}$. Formally, the embeddings for user $u$ and item $i$ in the $l$-th layer are:
\begin{equation}
\begin{aligned}
\mathbf{\hat{e}}_{u}^{(l)}
&=\frac{1}{\mathcal{N}(u)}
  \sum_{j \mid(u, j) \in \mathcal{E}}
  \frac{1}{\mathcal{N}(j)} \mathbf{\hat{e}}_{j}^{(l-1)},\\
\mathbf{\hat{e}}_{i}^{(l)}
&=\frac{1}{\mathcal{N}(i)}
  \sum_{v \mid(i, v) \in \mathcal{E}}
  \frac{1}{\mathcal{N}(v)} \mathbf{\hat{e}}_{v}^{(l-1)}.
\end{aligned}
\end{equation}
The initial enhanced user and item representations are:
\begin{equation}
\begin{aligned}
\mathbf{\hat{e}}_{u}^{(0)}
&=(\mathbf{e}_{u_t}|\mathbf{e}_{u_v}|t_{\theta}((P_u^t|P_u^v|P_u^m))),\\
\mathbf{\hat{e}}_{i}^{(0)}
&=(\mathbf{e}_{i_t}|\mathbf{e}_{i_v}|t_{\theta}((P_i^t|P_i^v|P_i^m))).
\end{aligned}
\end{equation}
Here $|$ denotes the concatenation operation. After $L$ layers of graph convolution, the final representations are:
\begin{equation}
\begin{aligned}
\mathbf{\hat{e}}_{u}&=\sum_{l=0}^L \mathbf{\hat{e}}_{u}^{(l)},&
\mathbf{\hat{e}}_{i}&=\sum_{l=0}^L \mathbf{\hat{e}}_{i}^{(l)}.
\end{aligned}
\end{equation}

Here, we fix $L=3$ for all experiments, which is the best setting in most multimodal recommendation models \citep{xu2025survey}. Entire user and item representations can be formulated as $\mathbf{\hat{E}}_{u}$ and $\mathbf{\hat{E}}_{i}$, respectively. Furthermore, we adopt constructed behavior- and multimodal-aware homogeneous graphs and refined unified item-item graph to enhance user and item representations. 

For the user side, we only have the constructed behavior- and multimodal-aware user-user graph $\mathcal{E}^{bmu}$ with similarity matrix $\mathcal{S}^{bmu}$. Therefore, user side representation enhancement can be expressed as:
\begin{equation}
    \mathbf{\bar{E}}_{u} = \mathbf{\hat{E}}_{u} +  \mathbf{\hat{E}}_{u}((\mathcal{D}^{bmu})^{-\frac{1}{2}} \mathcal{S}^{bmu}(\mathcal{D}^{bmu})^{-\frac{1}{2}}),
\end{equation}
where $\mathcal{D}^{bmu}$ is the diagonal degree matrix of $\mathcal{S}^{bmu}$. This normalization aims to mitigate the issues of gradient explosion or vanishing.

For the item side, we have the constructed behavior- and multimodal-aware item-item graph $\mathcal{E}^{bmi}$ with similarity matrix $\mathcal{S}^{bmi}$ and the refined unified item-item graph $\mathcal{E}^{rti}$ with similarity matrix $\dagger \mathcal{S}^{rti}$. Thus, item side representation enhancement can be expressed as:
\begin{equation}
\begin{aligned}
    \mathbf{\bar{E}}_{i}
    ={}& \mathbf{\hat{E}}_{i}
    + \mathbf{\hat{E}}_{i}\bigl((\mathcal{D}^{bmi})^{-\frac{1}{2}}
    \mathcal{S}^{bmi}(\mathcal{D}^{bmi})^{-\frac{1}{2}}\bigr)\\
    &+ \mathbf{\hat{E}}_{i}\bigl((\mathcal{D}^{rti})^{-\frac{1}{2}}
    \dagger \mathcal{S}^{rti}(\mathcal{D}^{rti})^{-\frac{1}{2}}\bigr),
\end{aligned}
\end{equation}
where $\mathcal{D}^{bmi}$ and $\mathcal{D}^{rti}$ are the diagonal row-degree matrices of $\mathcal{S}^{bmi}$ and $\dagger \mathcal{S}^{rti}$, respectively. These normalizations also aim to mitigate the issues of gradient explosion or vanishing. For efficiency, all homogeneous graphs use a single graph-convolution layer.

Consistent with almost all existing multimodal recommendation studies \citep{xu2025survey,liu2024multimodal,zhou2023comprehensive}, we use BPR for model optimization. Specifically, we compute the inner product of user and item representations to calculate predicted scores and adopt the BPR loss function:
\begin{equation}
\begin{aligned}
\mathcal{L}_{b p r}
= \sum_{(u, p, n) \in \mathcal{D}}
-\log \sigma\bigl(
\mathbf{\bar{e}}_u^{\top} \mathbf{\bar{e}}_p
- \mathbf{\bar{e}}_u^{\top}\mathbf{\bar{e}}_n
\bigr),
\end{aligned}
\end{equation}
where $\sigma(\cdot)$ is the sigmoid function, and $p$ and $n$ denote positive and negative items for user $u$. AgentMMRec can also be combined with self-supervised tasks \citep{xu2025mentor,zhou2023bootstrap,wei2023multi}; the default model omits them for efficiency, and Section~\ref{sec:ssl} reports the compatibility analysis.

\subsection{Reranking Final Item List}
\label{sec:reranking}
For each user $u$, the backbone recommender first retrieves a candidate list $\mathcal{I}_u$ of size $M$, where $M\geq N$ in the top-$N$ evaluation. In our implementation, candidates are drawn from the saved top-50 backbone list, so the reranker operates on a larger candidate pool than the reported $N\in\{10,20\}$. The Utilizer Agent reranks this candidate list by combining $u$'s behavior- and multimodal-aware preferences with the properties of candidate items. The top-$N$ prefix of the reranked list is used to compute Recall@$N$ and NDCG@$N$, which is why reranking can affect both metrics when $M>N$. This process is guided by $\mathcal{P}_{rank}$:
\begin{equation}
\begin{aligned}
    \bar{\mathcal{I}}_u
    \leftarrow \operatorname{UAgent}\bigl(&u,\mathcal{I}_u,
    (P_u^t|P_u^v|P_u^m),\\
    &\{(P_i^t|P_i^v|P_i^m)\}_{i \in \mathcal{I}_u}
    \mid \mathcal{P}_{rank}\bigr),
\end{aligned}
\end{equation}
where $\bar{\mathcal{I}}_u$ is the reranked candidate list. During training, we define $\operatorname{Eval}(u,\mathcal{I}_u)$ as the per-user NDCG@$N$ on the supervision split available to the training procedure. If $\operatorname{Eval}(u,\bar{\mathcal{I}}_u)<\operatorname{Eval}(u,\mathcal{I}_u)$, the reranking step has hurt the training signal, and the Integrator Agent updates $u$'s memory from the available ground-truth training items $\hat{\mathcal{I}}_u$:
\begin{equation}
\begin{aligned}
    (P_u^t|P_u^v|P_u^m)
    \leftarrow \operatorname{IAgent}\bigl(&u,\hat{\mathcal{I}}_u,
    (P_u^t|P_u^v|P_u^m),\\
    &\{(P_i^t|P_i^v|P_i^m)\}_{i \in \hat{\mathcal{I}}_u}
    \mid \mathcal{P}_{update}\bigr).
\end{aligned}
\end{equation}
After updating $u$'s preferences, the model reranks the candidate list again. The loop stops when reranking no longer hurts the training feedback or when the configured update budget is reached. For efficiency, updates are attempted every $E$ epochs. At validation and test time, the learned knowledge memory is frozen; no validation or test ground-truth items are used to update memory, construct graphs, revise prompts, or decide whether a reranked list is beneficial. The templates $\mathcal{P}_{rank}$ and $\mathcal{P}_{update}$ are included with the code (The code will be made publicly available upon acceptance of the paper).

\subsection{Algorithm Overview}
Algorithm~\ref{algorithm:process} summarizes the training and inference protocol.

\begin{algorithm*}[t]
\caption{Training and inference process of AgentMMRec}
\label{algorithm:process}
\vskip -0.1in
\centering
\footnotesize
\setlength{\tabcolsep}{4pt}
\renewcommand{\arraystretch}{1.12}
\begin{tabular}{@{}p{0.06\textwidth}p{0.18\textwidth}p{0.72\textwidth}@{}}
\toprule
\multicolumn{3}{@{}p{0.96\textwidth}@{}}{\textbf{Input:} split graphs $\mathcal{G}_{train},\mathcal{G}_{val},\mathcal{G}_{test}$; item text and image fields; encoders $t_\theta(\cdot),v_\theta(\cdot)$; agents $\operatorname{IAgent}(\cdot),\operatorname{UAgent}(\cdot)$; knowledge memory $\operatorname{KMemory}(\cdot)$; prompts; update interval $E$ and budget $B$.} \\
\multicolumn{3}{@{}p{0.96\textwidth}@{}}{\textbf{Output:} frozen knowledge memory and test recommendation lists.} \\
\midrule
\rowcolor{Gray}
\multicolumn{3}{@{}l}{\textbf{Phase I: Knowledge extraction and graph construction}} \\
\textbf{1} & Encode items &
Obtain $\mathbf{E}_{i_t}$ and $\mathbf{E}_{i_v}$ with $t_\theta(\cdot)$ and $v_\theta(\cdot)$; initialize $\mathbf{E}_{u_t}$ and $\mathbf{E}_{u_v}$. \\
\textbf{2} & User memory &
For each user $u$, extract $(P^{t}_u,P^{v}_u,P^{m}_u)$ from training items $\{T_i,V_i\}_{i\in\mathcal{N}_{train}(u)}$ and write the preferences into $\operatorname{KMemory}$. \\
\textbf{3} & User graph &
Encode stored user preferences and construct the behavior- and multimodal-aware user-user graph $\mathcal{G}^{bmu}$. \\
\textbf{4} & Item memory &
For each item $i$, extract $(P^{t}_i,P^{v}_i,P^{m}_i)$ from training co-purchased items $\{T_j,V_j\}_{j\in\mathcal{N}^{2}_{train}(i)}$ and write the properties into $\operatorname{KMemory}$. \\
\textbf{5} & Item graphs &
Construct $\mathcal{G}^{bmi}$ from stored item properties; refine modality-specific item graphs $\{\dagger\mathcal{S}^{i,m}\}_{m\in\{t,v\}}$ into $\mathcal{G}^{rti}$ with $\operatorname{UAgent}$. \\
\addlinespace[2pt]
\rowcolor{Gray}
\multicolumn{3}{@{}l}{\textbf{Phase II: Training with knowledge-enhanced reranking}} \\
\textbf{6} & Optimize &
Train the backbone recommender with knowledge-enhanced user/item representations and BPR loss until convergence. \\
\textbf{7} & Update memory &
Every $E$ epochs, rerank each candidate list with $\operatorname{UAgent}$. If reranking lowers training feedback, update that user's memory from training ground-truth items only, stopping when feedback recovers or the budget $B$ is reached. \\
\addlinespace[2pt]
\rowcolor{Gray}
\multicolumn{3}{@{}l}{\textbf{Phase III: Frozen validation and testing}} \\
\textbf{8} & Select &
Freeze $\operatorname{KMemory}$ and select hyperparameters on $\mathcal{G}_{val}$ without memory updates. \\
\textbf{9} & Test &
Generate and rerank test candidates on $\mathcal{G}_{test}$ using the frozen $\operatorname{KMemory}$ only. \\
\bottomrule
\end{tabular}
\vspace{-0.08in}
\end{algorithm*}

\section{Experiment}
\label{sec:experiment}
\subsection{Experiment Setup}

\subsubsection{Datasets}
The experiments are conducted on three real-world Amazon datasets with text and image modalities: Baby, Sports, and Clothing \citep{mcauley2015image}. The textual features are derived from item descriptions, and the visual features are derived from product images. We follow the MMRec preprocessing protocol \citep{zhou2023mmrecsm} and apply 5-core filtering to remove infrequent users and items. Table~\ref{tab:dataset_statistics} summarizes the filtered datasets. The processed interactions are split into training, validation, and test sets in an 8:1:1 ratio.

\begin{table}[!t]
    \centering
\caption{Statistics of all evaluation datasets.}
\vskip -0.1in
\label{tab:dataset_statistics}
    \begin{tabular}{ccccc}
    \toprule
         Datasets&  \#Users&  \#Items& \#Interactions& Sparsity\\
         \midrule
         Baby & 19,445 & 7,050 & 160,792 & 99.88\% \\
         Sports & 35,598 & 18,357 & 296,337 & 99.95\% \\
         Clothing & 39,387 & 23,033 & 278,677 & 99.97\% \\
         \bottomrule
    \end{tabular}
     \vskip -0.1in
\end{table}

\subsubsection{Evaluation Metrics}
Following previous multimodal recommendation studies \citep{xu2025survey,zhou2023bootstrap,zhou2023tale}, we report Recall@$N$ and NDCG@$N$ for $N=10$ and $N=20$. For reranking, the backbone first retrieves a candidate list of size $M=50$, and the Utilizer Agent reorders this list before the top-$N$ prefix is evaluated. All reported test metrics are computed after freezing the knowledge memory and model parameters. Significance markers in Table~\ref{tab:result} compare AgentMMRec with the strongest baseline under frozen test rankings.

\subsubsection{Baselines}
We compare AgentMMRec with 17 representative multimodal recommendation baselines. The main paper lists the baselines by family, with method-level descriptions provided in Appendix~\ref{appendix:baselines}. Graph- and fusion-based baselines include MMGCN~\citep{wei2019mmgcn}, DualGNN~\citep{wang2021dualgnn}, LATTICE~\citep{zhang2021mining}, FREEDOM~\citep{zhou2023tale}, LGMRec~\citep{guo2024lgmrec}, COHESION~\citep{xu2025cohesion}, and HPMRec~\citep{chen2025hypercomplex}. SSL- or denoising-oriented baselines include SLMRec~\citep{tao2022self}, BM3~\citep{zhou2023bootstrap}, MMSSL~\citep{wei2023multi}, DiffMM~\citep{jiang2024diffmm}, SMORE~\citep{ong2025spectrum}, BeFA~\citep{fan2025befa}, MENTOR~\citep{xu2025mentor}, EVEN~\citep{qi2025seeing}, and FreRec~\citep{peng2025frequency}. We also include LLMRec~\citep{wei2024llmrec} as the LLM-enhanced baseline.

\subsubsection{Implementation Details}
We keep the standard settings of the baselines and fix the batch size at 2048. Hyperparameters are tuned according to the optimal configurations reported in the corresponding papers. All baselines are implemented in PyTorch with Adam \citep{kingma2014adam} and Xavier initialization \citep{glorot2010understanding}. To isolate the effect of the agent-memory design from stronger encoders, AgentMMRec uses the same MMRec text and vision encoders $t_\theta(\cdot)$ and $v_\theta(\cdot)$ as the baselines \citep{xu2025survey}. The default Integrator and Utilizer backbone is Qwen2.5-VL-7B; Appendix~\ref{appendix:backbone} studies Qwen2.5-VL-32B and GPT-4o-2024-08-06. We set $k=20$, $\Upsilon=5$, the memory update interval to $E=10$, the learning rate to $10^{-3}$, and the early-stopping patience to 20 validation checks. The LLM decoding temperature is 0.7 with a maximum output length of 2,000 tokens. All results are from the mean of 5 runs of random seeds.

\subsubsection{Leakage Control}
All preference extraction, property extraction, graph construction, graph refinement, reranking feedback, and memory updates use only the training split. The validation split is used for model selection and hyperparameter tuning after the memory has been constructed from training data. During test evaluation, the knowledge memory and homogeneous graphs are frozen, and the Utilizer Agent receives only the candidate items and the frozen preference/property entries. Test interactions are never used to update memory, construct graphs, revise prompts, or select reranking updates. Algorithm~\ref{algorithm:process} summarizes the same protocol across training, validation, and testing.

\begin{table*}[!t]
    \centering
     \caption{Performance comparison of baselines and AgentMMRec in terms of Recall and NDCG. $^*$ indicates that the t-tests validate the significance of performance improvements with $p$-value $<$ 0.05.}
     \vskip -0.1in
    \label{tab:result}
\resizebox{\linewidth}{!}{
    \begin{tabular}{l|cccccccccccc}
    \toprule
          \multicolumn{1}{c}{Datasets}&  \multicolumn{4}{c}{Baby}&  \multicolumn{4}{c}{Sports}&  \multicolumn{4}{c}{Clothing}\\\midrule
          \multicolumn{1}{c}{Metrics}& R@10& R@20& N@10& N@20& R@10& R@20& N@10& N@20& R@10& R@20& N@10& N@20 \\\midrule
          MMGCN (MM'19)
& 0.0378 &0.0615& 0.0200 &0.0261& 0.0370 &0.0605& 0.0193 &0.0254 &0.0218& 0.0345& 0.0110& 0.0142\\
          DualGNN (TMM'21) 
& 0.0448 &0.0716& 0.0240 &0.0309& 0.0568 &0.0859& 0.0310 &0.0385 &0.0454 &0.0683 &0.0241 &0.0299\\
          LATTICE (MM'21)
& 0.0547 &0.0850& 0.0292 &0.0370& 0.0620 &0.0953& 0.0335 &0.0421 &0.0492 &0.0733 &0.0268 &0.0330\\
          SLMRec (TMM'22)
& 0.0529 &0.0775& 0.0290 &0.0353& 0.0663 &0.0990& 0.0365 &0.0450 &0.0452 &0.0675 &0.0247 &0.0303\\
          FREEDOM (MM'23)
& 0.0627 &0.0992& 0.0330 & 0.0424& 0.0717&0.1089& 0.0385 &0.0481 &0.0629 &0.0941 &0.0341 &0.0420\\
          BM3 (WWW'23)
& 0.0564 &0.0883& 0.0301 &0.0383& 0.0656 &0.0980& 0.0355 &0.0438 &0.0422 &0.0621 &0.0231 &0.0281\\
          MMSSL (WWW'23)
& 0.0613 &0.0971& 0.0326 &0.0420& 0.0693 &0.1013& 0.0369 &0.0474 &0.0531 &0.0797 &0.0291 &0.0359\\
          LLMRec (WSDM'24)
& 0.0621 & 0.0983 & 0.0324 & 0.0422 & 0.0682 & 0.1000 & 0.0363 & 0.0459 & 0.0540& 0.0808& 0.0294& 0.0365\\
          LGMRec (AAAI'24) 
& 0.0639& 0.0989& 0.0337& 0.0430& 0.0719& 0.1068& 0.0387& 0.0477& 0.0555& 0.0828& 0.0302& 0.0371\\
          DiffMM (MM'24)
& 0.0623& 0.0975& 0.0328& 0.0411& 0.0671& 0.1017& 0.0377& 0.0458& 0.0531& 0.0797& 0.0291& 0.0359\\
          SMORE (WSDM'25)
& \underline{0.0680}& 0.1035& \underline{0.0365}& \underline{0.0457}& 0.0762& 0.1142& 0.0408& 0.0506& 0.0659& 0.0987& \underline{0.0360}& 0.0443\\
          FreRec (MM'25)& 0.0662& 0.1011& 0.0348& 0.0437& 0.0754& \underline{0.1147}& 0.0410& 0.0508& 0.0674& 0.0977& \underline{0.0363}& \underline{0.0447}\\
          EVEN (AAAI'25)& 0.0667& 0.1031& 0.0355& 0.0448& 0.0759& 0.1143& \underline{0.0411}& 0.0510& 0.0662& 0.0978& 0.0356& 0.0436\\
          BeFA (AAAI'25)
& 0.0555& 0.0884& 0.0299& 0.0383& 0.0649& 0.0985& 0.0346& 0.0432& 0.0568& 0.0857& 0.0307& 0.0381\\ 
          MENTOR (AAAI'25)
& 0.0678& 0.1048& 0.0362& 0.0450& \underline{0.0763}& 0.1139& 0.0409& \underline{0.0511}& \underline{0.0668}& \underline{0.0989}& 0.0360& 0.0441\\
          COHESION (SIGIR'25)
& \underline{0.0680}& \underline{0.1052}& 0.0354& 0.0454& 0.0752& 0.1137& 0.0409& 0.0503& 0.0665& 0.0983& 0.0358& 0.0438\\
          HPMRec (CIKM'25)
& 0.0667& 0.1033& 0.0357& 0.0451& 0.0751& 0.1129& 0.0410& 0.0507& 0.0658& 0.0963& 0.0351& 0.0429\\
\midrule
          AgentMMRec (Qwen)
& \textbf{0.0705$^*$}& \textbf{0.1079$^*$}& \textbf{0.0380$^*$}& \textbf{0.0475$^*$}& \textbf{0.0838$^*$}& \textbf{0.1231$^*$}& \textbf{0.0454$^*$}& \textbf{0.0557$^*$}& \textbf{0.0740$^*$}& \textbf{0.1071$^*$}& \textbf{0.0404$^*$}& \textbf{0.0490$^*$}\\
\bottomrule
\end{tabular}
    }
    \vskip -0.1in
\end{table*}

\subsection{Overall Performance}
We evaluate the effectiveness of AgentMMRec on multiple real-world datasets in multimodal recommendation scenarios. From Table~\ref{tab:result}, we find the following observations:
\begin{itemize}[leftmargin=*]
    \item AgentMMRec improves over all baselines across datasets. The result supports the central design choice: user and item knowledge extracted from multimodal content becomes more useful when it is injected into graph construction, representation learning, and reranking rather than used only as augmented text.
    \item Strong baselines such as SMORE, MENTOR, COHESION, and HPMRec remain close to one another despite different architectures. This pattern suggests that feature-level multimodal fusion faces a shared bottleneck. Section~\ref{sec:compatibility} tests this explanation by transferring AgentMMRec's behavior-aware graphs and reranking memory to these backbones.
\end{itemize}

\begin{figure}[!t]
    \centering
    \includegraphics[width=1\linewidth]{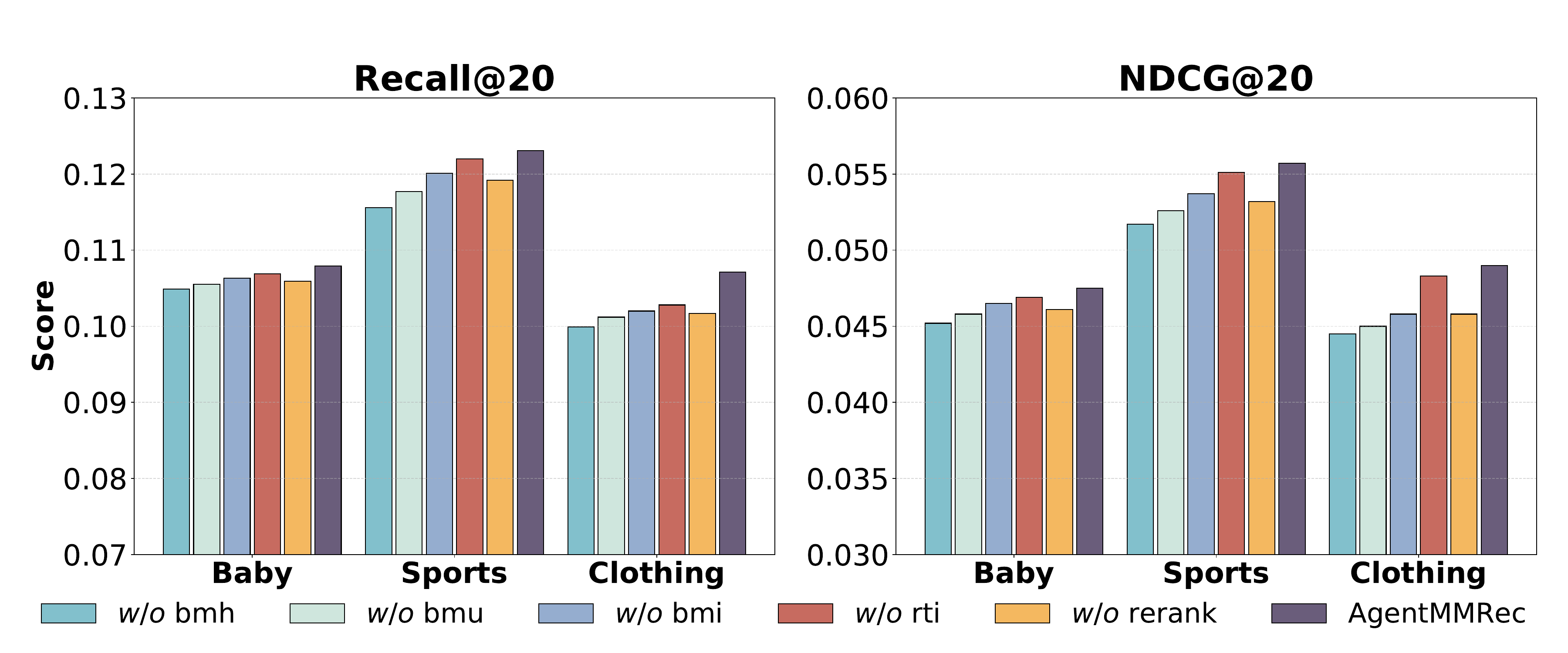}
    \vskip -0.1in
    \caption{Ablation study across all datasets.}
    \label{fig:ablation}
     \vskip -0.1in
\end{figure}

\begin{table*}[!t]
    \centering
     \caption{Compatibility analysis of AgentMMRec with suboptimal baselines.}
     \small
       \vskip -0.1in
    \label{tab:compatibility}
    \begin{tabular}{ll|cccccccccccc}
    \toprule
          \multirow{2.5}{*}{Models}& \multicolumn{1}{c}{Datasets}&  \multicolumn{4}{c}{Baby}&  \multicolumn{4}{c}{Sports}&  \multicolumn{4}{c}{Clothing}\\\cmidrule{2-14}
          &\multicolumn{1}{c}{Metrics}& R@10& R@20& N@10& N@20& R@10& R@20& N@10& N@20& R@10& R@20& N@10& N@20 \\
         
          \midrule SMORE& Original
& 0.0680& 0.1035& 0.0365& 0.0457& 0.0762& 0.1142& 0.0408& 0.0506& 0.0659& 0.0987& 0.0360& 0.0443\\
                        & $+Graph$
& \textbf{0.0691}& \textbf{0.1055}& \textbf{0.0371}& \textbf{0.0466}& \textbf{0.0799}& \textbf{0.1190}& \textbf{0.0437}& \textbf{0.0532}& \textbf{0.0709}& \textbf{0.1041}& \textbf{0.0388}& \textbf{0.0473}\\
                        & $+Rerank$
& \underline{0.0686}& \underline{0.1047}& \underline{0.0369}& \underline{0.0463}& \underline{0.0785}& \underline{0.1170}& \underline{0.0431}& \underline{0.0527}& \underline{0.0688}& \underline{0.1021}& \underline{0.0377}& \underline{0.0462}\\
         
          \midrule MENTOR& Original
& 0.0678& 0.1048& 0.0362& 0.0450& 0.0763& 0.1139& 0.0409& 0.0511& 0.0668& 0.0989& 0.0360& 0.0441\\
                         & $+Graph$
& \textbf{0.0693}& \textbf{0.1061}& \textbf{0.0370}& \textbf{0.0461}& \textbf{0.0792}& \textbf{0.1180}& \textbf{0.0434}& \textbf{0.0532}& \textbf{0.0707}& \textbf{0.1035}& \textbf{0.0383}& \textbf{0.0466}\\
                         & $+Rerank$
& \underline{0.0685}& \underline{0.1053}& \underline{0.0366}& \underline{0.0453}& \underline{0.0778}& \underline{0.1164}& \underline{0.0427}& \underline{0.0528}& \underline{0.0689}& \underline{0.1024}& \underline{0.0370}& \underline{0.0455}\\

          \midrule COHESION& Original
& 0.0680& 0.1052& 0.0354& 0.0454& 0.0752& 0.1137& 0.0409& 0.0503& 0.0665& 0.0983& 0.0358& 0.0438\\
                           & $+Graph$
& \textbf{0.0695}& \textbf{0.1066}& \textbf{0.0365}& \textbf{0.0460}& \textbf{0.0780}& \textbf{0.1174}& \textbf{0.0430}& \textbf{0.0525}& \textbf{0.0697}& \textbf{0.1033}& \textbf{0.0380}& \textbf{0.0462}\\
                           & $+Rerank$
& \underline{0.0686}& \underline{0.1059}& \underline{0.0358}& \underline{0.0458}& \underline{0.0773}& \underline{0.1159}& \underline{0.0423}& \underline{0.0519}& \underline{0.0681}& \underline{0.1015}& \underline{0.0370}& \underline{0.0453}\\

          \midrule HPMRec& Original
& 0.0667& 0.1033& 0.0357& 0.0451& 0.0751& 0.1129& 0.0410& 0.0507& 0.0658& 0.0963& 0.0351& 0.0429\\
                         & $+Graph$
& \textbf{0.0682}& \textbf{0.1054}& \textbf{0.0366}& \textbf{0.0459}& \textbf{0.0785}& \textbf{0.1174}& \textbf{0.0432}& \textbf{0.0530}& \textbf{0.0698}& \textbf{0.1025}& \textbf{0.0375}& \textbf{0.0449}\\
                         & $+Rerank$
& \underline{0.0677}& \underline{0.1042}& \underline{0.0360}& \underline{0.0454}& \underline{0.0776}& \underline{0.1162}& \underline{0.0427}& \underline{0.0525}& \underline{0.0684}& \underline{0.1004}& \underline{0.0362}& \underline{0.0440}\\

\bottomrule
\end{tabular}
    \vskip -0.1in
\end{table*}

\begin{figure}[!t]
    \centering
    \includegraphics[width=1\linewidth]{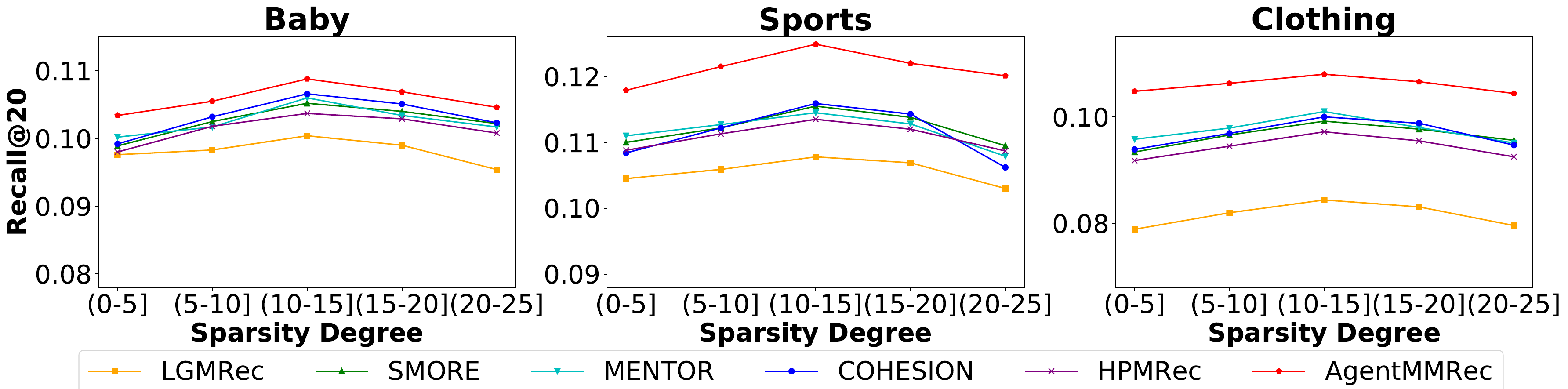}
     \vskip -0.15in
     \caption{Sparsity analysis across all datasets.}
    \label{fig:sparsity}
     \vskip -0.15in
\end{figure}

\subsection{Ablation Study}
\label{sec:ablation}
To validate the effectiveness of AgentMMRec, we conduct experiments to justify the importance of key components. We design the following variants: (1) $w/o$ bmh, which removes both behavior- and multimodal-aware user-user and item-item graphs. (2) $w/o$ bmu, which removes the behavior- and multimodal-aware user-user graph. (3) $w/o$ bmi, which removes the behavior- and multimodal-aware item-item graph. (4) $w/o$ rti, which directly uses traditional item-item graphs to replace the unified item-item graph. (5) $w/o$ rerank, which removes the rerank and feedback process. Notably, for variants (1)-(3), Integrator Agent still extracts and stores behavior- and multimodal-aware user preferences and item properties.

Figure~\ref{fig:ablation} shows that each component contributes to the performance improvement of AgentMMRec. We provide the following in-depth observations:
\begin{itemize}[leftmargin=*]
    \item \textbf{Behavior-aware homogeneous graphs.} Removing both behavior- and multimodal-aware graphs leads to consistent degradation, showing that the extracted knowledge contributes beyond the LLM text itself. The drop is larger on Sports and Clothing, where sparser interactions make supplementary graph connectivity more valuable.
    \item \textbf{User-user and item-item graphs.} The item-item graph contributes more on Sports and Clothing, while the user-user graph is relatively more useful on Baby. This is consistent with the dataset structure: Baby has fewer items and denser user interactions, whereas the larger Sports and Clothing catalogs benefit more from item-property relations.
    \item \textbf{Graph refinement.} Replacing the refined unified item-item graph with traditional modality-specific graphs hurts performance across datasets. The result indicates that content-only neighbors are insufficient when modality similarity conflicts with behavior-aware item relevance.
    \item \textbf{Reranking and feedback.} Removing reranking and feedback yields a smaller but consistent drop. The result suggests that the main gains come from graph and representation enhancement, while reranking provides an additional late-stage adjustment.
\end{itemize}
In Section~\ref{sec:compatibility}, we further explore whether the key components of AgentMMRec can be transferred to existing models to break through their performance bottlenecks.

\subsection{Compatibility Analysis}
\label{sec:compatibility}
We conduct two compatibility experiments to test whether AgentMMRec's constructed knowledge benefits existing models. The +$Graph$ variant transfers AgentMMRec's behavior- and multimodal-aware homogeneous graphs to selected baselines. The +$Rerank$ variant lets the Utilizer Agent rerank each baseline's output with the knowledge memory optimized by AgentMMRec. We select SMORE, MENTOR, COHESION, and HPMRec from Table~\ref{tab:result}. For +$Graph$, we follow previous studies~\citep{zhou2023tale,xu2025cohesion} and run graph convolution with the transferred graphs.

Results in Table~\ref{tab:compatibility} lead to several important findings:
\begin{itemize}[leftmargin=*]
    \item \textbf{Graph transfer.} The +$Graph$ variant improves all four baselines on all datasets. The gains are largest on Sports, where Recall@20 increases by 4.2\% for SMORE, 3.6\% for MENTOR, 3.3\% for COHESION, and 4.0\% for HPMRec. This pattern supports the claim that behavior-aware graph structure addresses a bottleneck shared by different multimodal backbones.
    \item \textbf{Knowledge-enhanced reranking.} The +$Rerank$ variant also improves the baselines, but usually less than graph transfer because it changes only the final candidate order. The result still shows that the memory contains user preference and item property information that complements the baseline rankings.
    \item \textbf{Transfer across architectures.} SMORE, MENTOR, COHESION, and HPMRec use different fusion mechanisms, yet all benefit from AgentMMRec's knowledge. The compatibility result indicates that the proposed memory is not tied to a single backbone implementation.
\end{itemize}

\subsection{Sparsity Analysis}
\label{sec:sparsity_analysis}
We evaluate AgentMMRec under different user sparsity levels by grouping users according to their number of training interactions. We compare against LGMRec, SMORE, MENTOR, COHESION, and HPMRec on the derived sub-datasets.

Figure~\ref{fig:sparsity} shows that AgentMMRec outperforms these baselines across sparsity groups:
\begin{itemize}[leftmargin=*]
    \item \textbf{Sparse users.} The largest gains appear in $0$--$5$ interaction group, where collaborative evidence is weakest and behavior-aware knowledge extracted from few observed items is most useful.
    \item \textbf{Denser users.} The gains narrow as interaction counts increase, but remain positive in the $20$--$25$ group. This suggests that the graph and memory components add information beyond compensating for user sparsity.
    \item \textbf{Dataset differences.} Sports and Clothing have larger item catalogs and higher sparsity than Baby, and the item-side graph contributes more on these datasets. This is consistent with the ablation trend in Figure~\ref{fig:ablation}.
\end{itemize}

\begin{table*}[!t]
    \centering
    \caption{Item cold-start analysis across all datasets.}
    \label{tab:cold-start}
     \vskip -0.1in
     \small
    \begin{tabular}{l|cccccccccccc}
    \toprule
          \multicolumn{1}{c}{Datasets}&  \multicolumn{4}{c}{Baby}&  \multicolumn{4}{c}{Sports}&  \multicolumn{4}{c}{Clothing}\\\midrule
          \multicolumn{1}{c}{Metrics}& R@10& R@20& N@10& N@20& R@10& R@20& N@10& N@20& R@10& R@20& N@10& N@20 \\\midrule
          MMGCN
& 0.0103& 0.0186& 0.0062& 0.0098& 0.0106& 0.0178& 0.0060& 0.0094& 0.0069& 0.0101& 0.0039& 0.0050\\
          DualGNN
& 0.0132& 0.0200& 0.0077& 0.0110& 0.0166& 0.0242& 0.0096& 0.0132& 0.0134& 0.0198& 0.0082& 0.0113\\
          LATTICE
& 0.0175& 0.0256& 0.0099& 0.0138& 0.0266& 0.0340& 0.0139& 0.0189& 0.0168& 0.0249& 0.0095& 0.0135\\
          SLMRec
& 0.0172& 0.0259& 0.0101& 0.0140& 0.0281& 0.0354& 0.0142& 0.0194& 0.0141& 0.0208& 0.0084& 0.0118\\
          FREEDOM
& 0.0348& 0.0588& 0.0195& 0.0257& 0.0389& 0.0640& 0.0231& 0.0289& 0.0339& 0.0585& 0.0190& 0.0252\\
          BM3
& 0.0180& 0.0262& 0.0100& 0.0133& 0.0210& 0.0294& 0.0128& 0.0180& 0.0125& 0.0189& 0.0077& 0.0104\\
          MMSSL
& 0.0280& 0.0351& 0.0144& 0.0192& 0.0299& 0.0370& 0.0152& 0.0203& 0.0200& 0.0294& 0.0108& 0.0157\\
          LLMRec
& 0.0380& 0.0605& 0.0203& 0.0255& 0.0361& 0.0600& 0.0208& 0.0261& 0.0298& 0.0539& 0.0169& 0.0226\\
          LGMRec
& 0.0371& 0.0592& 0.0208& 0.0261& 0.0380& 0.0629& 0.0226& 0.0281& 0.0303& 0.0551& 0.0173& 0.0235\\
          DiffMM
& 0.0336& 0.0552& 0.0193& 0.0238& 0.0355& 0.0589& 0.0202& 0.0254& 0.0266& 0.0510& 0.0150& 0.0221\\
          SMORE
& 0.0370& 0.0595& 0.0202& 0.0251& 0.0404& 0.0661& 0.0245& \underline{0.0302}& 0.0360& 0.0602& 0.0195& 0.0259\\
          BeFA
& 0.0188& 0.0262& 0.0104& 0.0149& 0.0220& 0.0303& 0.0134& 0.0182& 0.0232& 0.0367& 0.0131& 0.0200\\
          MENTOR
& 0.0395& 0.0628& \underline{0.0212}& \underline{0.0268}& 0.0402& 0.0661& \underline{0.0249}& 0.0297& \underline{0.0369}& 0.0610& \underline{0.0201}& \underline{0.0264}\\
          COHESION
& \underline{0.0399}& \underline{0.0631}& 0.0211& 0.0263& \underline{0.0410}& \underline{0.0665}& 0.0246& 0.0300& \underline{0.0369}& \underline{0.0611}& 0.0199& 0.0256\\
          HPMRec
& 0.0378& 0.0603& 0.0204& 0.0256& 0.0398& 0.0653& 0.0241& 0.0292& 0.0360& 0.0600& 0.0192& 0.0255\\
\midrule
          AgentMMRec
& \textbf{0.0458}& \textbf{0.0733}& \textbf{0.0248}& \textbf{0.0317}& \textbf{0.0454}& \textbf{0.0711}& \textbf{0.0272}& \textbf{0.0324}& \textbf{0.0406}& \textbf{0.0662}& \textbf{0.0228}& \textbf{0.0282}\\
\bottomrule
\end{tabular}
    \vskip -0.15in
\end{table*}

\subsection{Cold-Start Analysis}
\label{sec:cold-start}
For item cold-start analysis, we follow prior settings \citep{zhang2022latent,xu2025survey}: 20\% of items are removed from training and split evenly into validation and test sets. The removed target items are excluded from the training interaction graph, while their text and image fields remain available as item content. As shown in Table~\ref{tab:cold-start}, AgentMMRec achieves the best performance across the three datasets. The gain indicates that behavior-aware knowledge helps align cold-start item content with the recommendation task when direct interaction evidence is limited. Additional analyses in Appendix~\ref{appendix:memory}, Appendix~\ref{appendix:backbone}, Appendix~\ref{appendix:ssl}, Appendix~\ref{appendix:efficiency}, and Appendix~\ref{appendix:template} cover memory updating, LLM backbone choice, SSL compatibility, efficiency and LLM cost, and template robustness.

\subsection{Hyperparameter Analysis}
\label{sec:hyperparameter}
\begin{figure}[!htbp]
    \centering
    \begin{subfigure}[b]{0.46\linewidth}
        \centering
        \includegraphics[width=\linewidth]{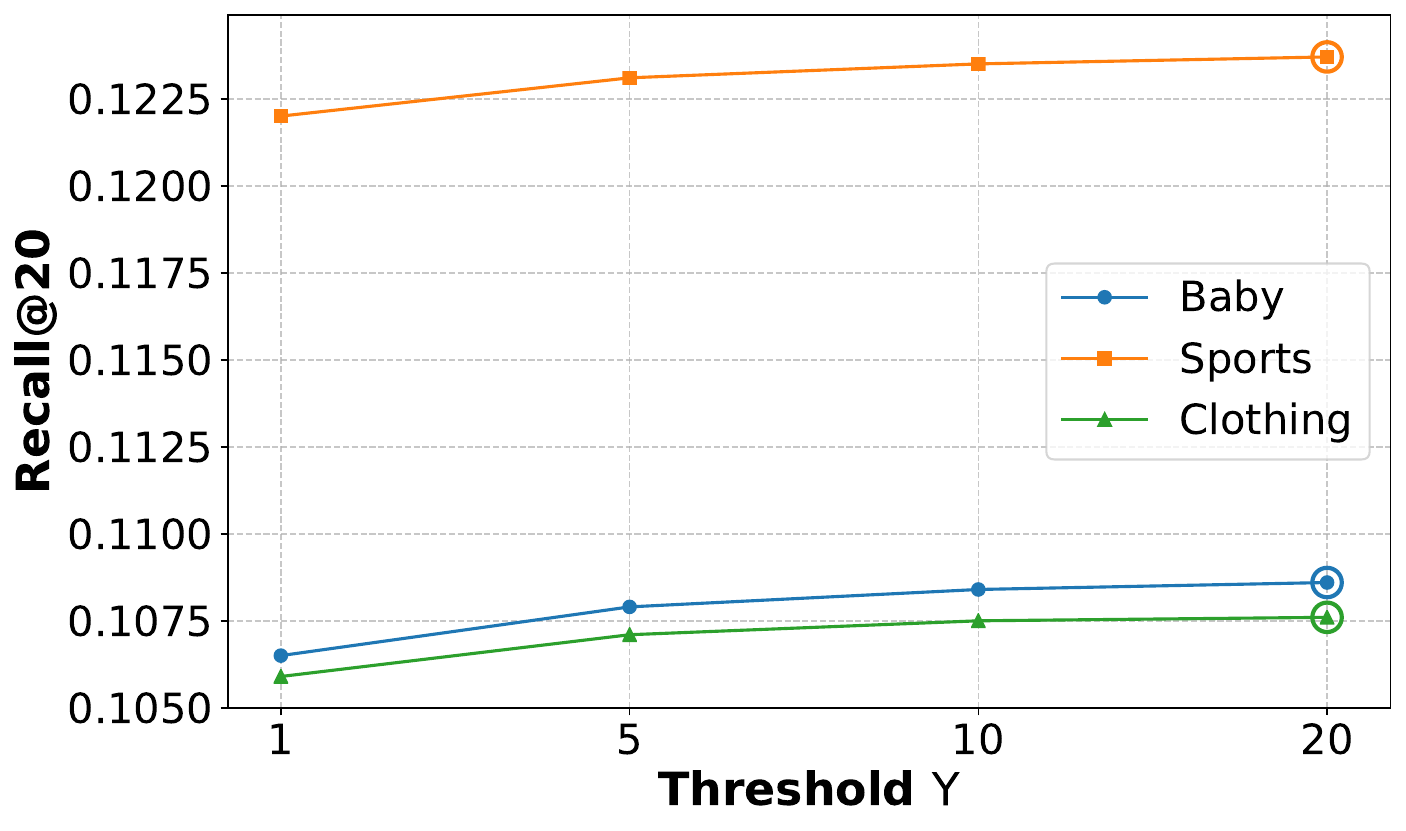}
        \caption{ThresholdR20}
    \end{subfigure}
    \hfill
    \begin{subfigure}[b]{0.46\linewidth}
        \centering
        \includegraphics[width=\linewidth]{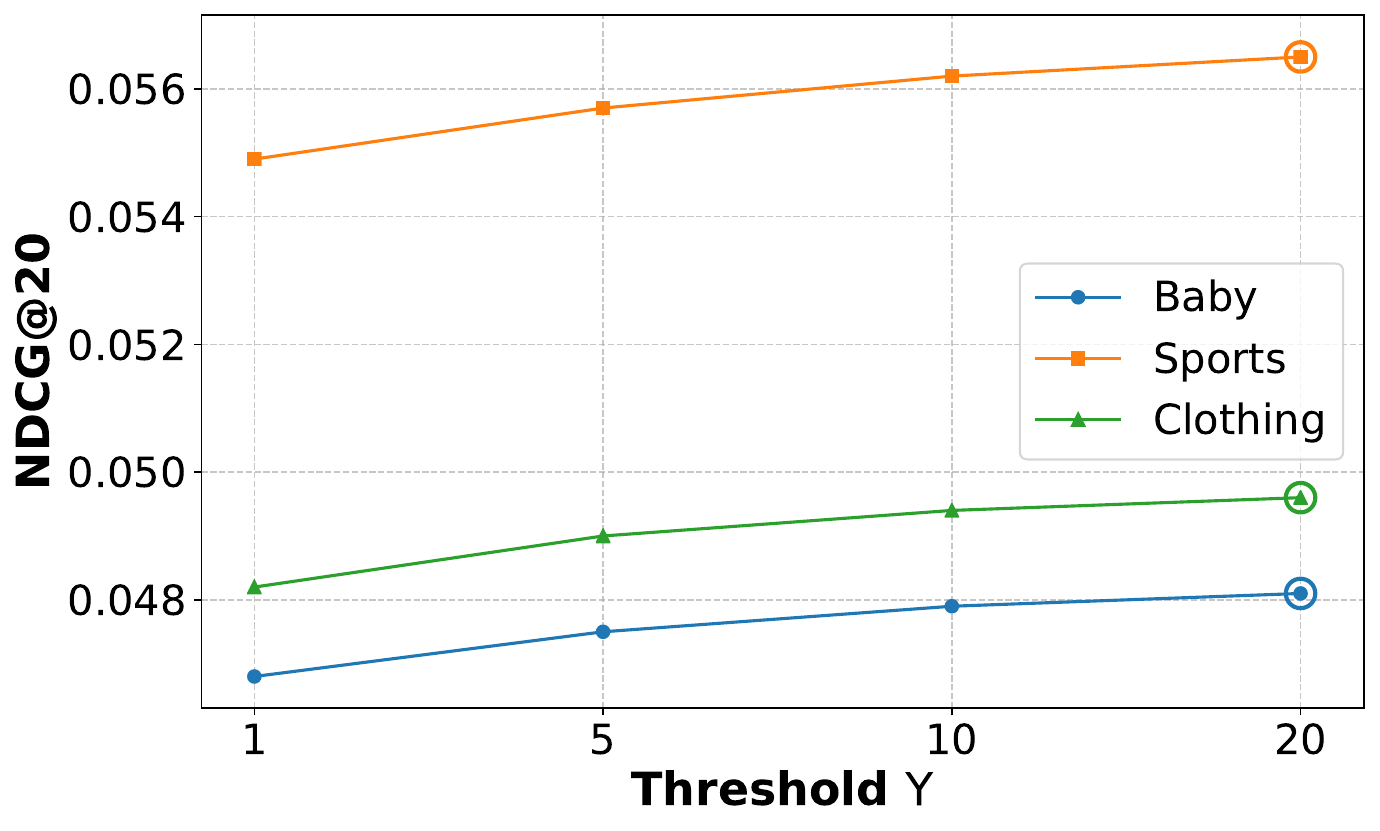}
        \caption{ThresholdN20}
    \end{subfigure}
    
    \begin{subfigure}[b]{0.46\linewidth}
        \centering
        \includegraphics[width=\linewidth]{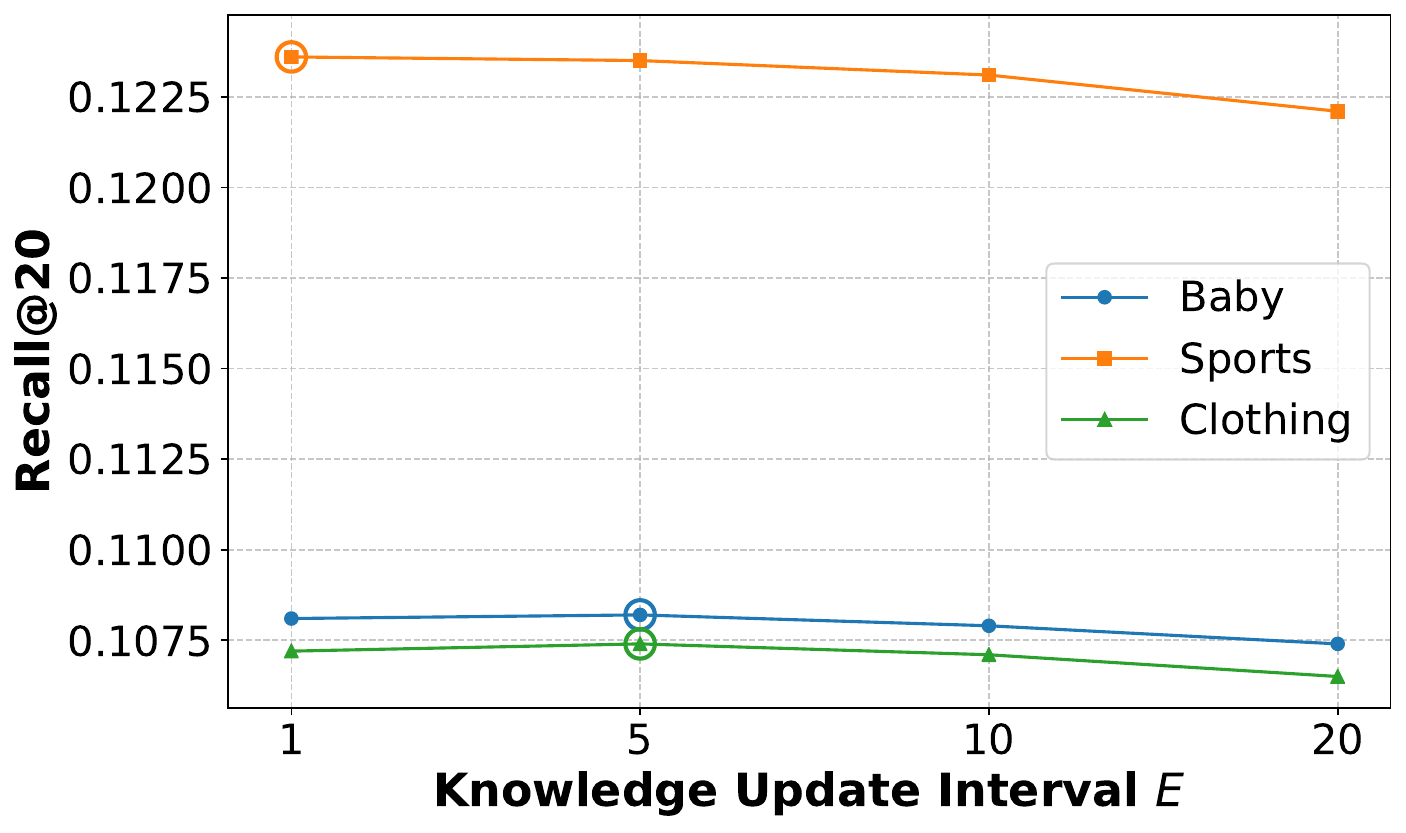}
        \caption{IntervalR20}
    \end{subfigure}
    \hfill
    \begin{subfigure}[b]{0.46\linewidth}
        \centering
        \includegraphics[width=\linewidth]{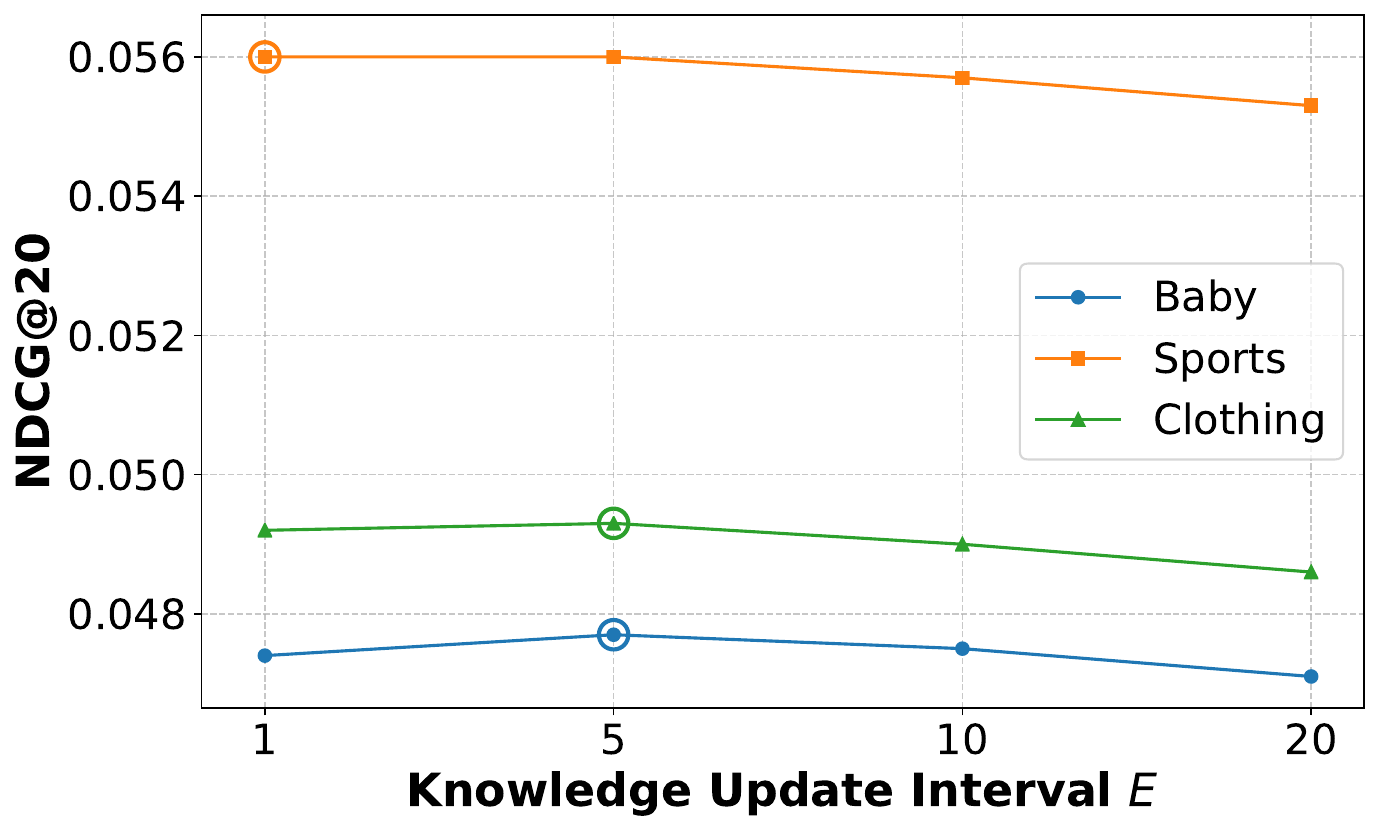}
        \caption{IntervalN20}
    \end{subfigure}
     \vskip -0.1in
    \caption{Effect of Threshold $\Upsilon$ and Knowledge Update Interval $E$.}
    \label{fig:hyperparameter}
     \vskip -0.15in
\end{figure}

We study the threshold $\Upsilon$ and the knowledge update interval $E$ in Figure~\ref{fig:hyperparameter}. The best result of each curve is marked.

Lower $\Upsilon$ reduces LLM cost but can sample less representative histories, while larger $\Upsilon$ uses more context and increases inference cost. Smaller $E$ updates the memory more frequently but can introduce noisier feedback and more LLM calls. We use $\Upsilon=5$ and $E=10$ in the main experiments to balance accuracy and cost rather than reporting only the best hyperparameter setting.

\subsection{Compatibility with Extra Modality-alignment SSL Tasks}
\label{sec:ssl}
We test whether AgentMMRec benefits from extra modality-alignment SSL. SSL is widely used in multimodal recommendation to align representations across modalities without additional labels \citep{wei2023multi,xu2025mentor,zhou2023bootstrap}. We incorporate two representative objectives: InfoNCE contrastive learning and distribution alignment (DisAlign).

Appendix~\ref{appendix:ssl} and Table~\ref{tab:SSL} show the detailed results:
\begin{itemize}[leftmargin=*]
    \item \textbf{Positive but limited gains.} InfoNCE and DisAlign consistently improve AgentMMRec, but the average gains are modest, about 0.3\%--0.5\% in Recall@20 and NDCG@20. This indicates that the agent-memory component is compatible with representation-level alignment, while much of the cross-modal signal may already be captured by behavior-aware knowledge extraction.
    \item \textbf{Objective differences.} InfoNCE is slightly better on Baby Recall@20, DisAlign is stronger on Clothing Recall@20, and the two objectives tie on Sports Recall@20. The differences are small, so we treat SSL as a complementary option.
    \item \textbf{Efficiency.} Because SSL objectives add augmentation, additional forward passes, and extra loss computation, the default AgentMMRec configuration omits them and uses the agent-memory design as the main source of multimodal alignment.
\end{itemize}

\subsection{Efficiency Study}
\label{sec:efficiency}
Appendix~\ref{appendix:efficiency} reports recommender training cost and the offline LLM call calculation. Since the agent-generated graphs are precomputed, AgentMMRec keeps the per-epoch recommender training loop close to standard graph-based multimodal recommenders.

Prompt-template robustness and the location of the full template files are described in Appendix~\ref{appendix:template}.

\section{Conclusion}
\label{sec:conclusion}
We presented AgentMMRec, a two-agent framework for bridging the semantic gap between multimodal item content and recommendation objectives. The Integrator Agent builds behavior-aware user and item knowledge from training interactions, and the Utilizer Agent converts this memory into refined graph structure, enhanced representations, and candidate reranking. The memory is frozen during validation and test evaluation, keeping ground-truth feedback out of the reported test results. Experiments on three multimodal recommendation datasets show consistent gains over recent baselines, with additional evidence from compatibility, sparsity, item cold-start, memory updating, backbone, SSL, template, and cost analyses. These results suggest that multimodal recommendation can benefit from treating LLM-derived semantics as reusable task-aware structure, not only as augmented features or final-stage ranking text.



\bibliographystyle{ACM-Reference-Format}
\balance

\newpage

\appendix
{\small
\section{Baseline Details}
\label{appendix:baselines}

\begin{sloppypar}
This section expands the compact baseline list in Section~\ref{sec:experiment}. We compare AgentMMRec with graph-based, self-supervised, denoising, fusion-oriented, and LLM-enhanced multimodal recommenders:

\begin{itemize}[leftmargin=*]
    \item \textbf{MMGCN} \citep{wei2019mmgcn}: MMGCN applies graph convolutional networks to model user-item interactions with modality-specific item features. It represents an early graph-based multimodal recommender that propagates visual and textual signals through the interaction graph.
    \item \textbf{DualGNN} \citep{wang2021dualgnn}: DualGNN jointly models user-item interactions and an auxiliary user-user graph. The user-user graph captures behavioral similarity among users and provides a reference point for evaluating whether AgentMMRec's behavior-aware user graph brings additional benefit.
    \item \textbf{LATTICE} \citep{zhang2021mining}: LATTICE constructs item-item semantic graphs from multimodal item features and injects the learned item relations into recommendation. It is a strong comparison for item graph construction because AgentMMRec also builds item-side structure but conditions it on stored user and item knowledge.
    \item \textbf{SLMRec} \citep{tao2022self}: SLMRec introduces self-supervised learning for multimodal recommendation through feature perturbation and modality-pattern discovery tasks. It evaluates whether auxiliary self-supervision alone can close the semantic gap between multimodal content and recommendation behavior.
    \item \textbf{FREEDOM} \citep{zhou2023tale}: FREEDOM denoises the user-item graph and uses a frozen item-item graph built from original modality features. It is relevant because it separates graph construction from online training and therefore provides a close comparison for precomputed item relations.
    \item \textbf{BM3} \citep{zhou2023bootstrap}: BM3 simplifies multimodal self-supervised learning by using dropout-based representation perturbation instead of heavy contrastive augmentations. It tests whether lightweight representation regularization can achieve the same gains as explicit knowledge memory.
    \item \textbf{MMSSL} \citep{wei2023multi}: MMSSL learns modality-aware interactive structures with adversarial perturbation and cross-modal contrastive learning. It distinguishes modality-common and modality-specific signals, making it a representative alignment-oriented baseline.
    \item \textbf{LLMRec} \citep{wei2024llmrec}: LLMRec uses LLM-based graph augmentation strategies to enrich recommendation signals. It is the closest baseline among the compared methods in terms of using LLM-derived knowledge, while AgentMMRec further stores the generated knowledge as reusable memory for graph construction, representation enhancement, and reranking.
    \item \textbf{LGMRec} \citep{guo2024lgmrec}: LGMRec combines local topological information and global embedding information through a hypergraph structure. It compares against AgentMMRec's homogeneous graph enhancement from the perspective of higher-order relation modeling.
    \item \textbf{DiffMM} \citep{jiang2024diffmm}: DiffMM integrates a modality-aware graph diffusion model with cross-modal contrastive learning. It is included to assess whether diffusion-based denoising and representation refinement can match the behavior-aware knowledge construction in AgentMMRec.
    \item \textbf{SMORE} \citep{ong2025spectrum}: SMORE reduces modality noise by exploiting discriminative spectrum properties and global frequency-domain information. It is a recent strong baseline for multimodal noise reduction and is also used in the compatibility analysis.
    \item \textbf{BeFA} \citep{fan2025befa}: BeFA corrects multimodal features according to user behavior. It provides a behavior-aware feature correction baseline, while AgentMMRec converts behavior-aware knowledge into both memory and graph structure.
    \item \textbf{MENTOR} \citep{xu2025mentor}: MENTOR designs multi-level cross-modal alignment tasks to improve the final user and item representations. It is a strong alignment baseline and tests whether explicit preference and property memory adds information beyond representation alignment.
    \item \textbf{COHESION} \citep{xu2025cohesion}: COHESION introduces a dual-stage fusion mechanism for composite graphs. It is included because it directly studies graph fusion quality, making it a natural comparison for AgentMMRec's behavior- and multimodal-aware graph construction.
    \item \textbf{HPMRec} \citep{chen2025hypercomplex}: HPMRec uses hypercomplex operations to enrich feature diversity and bridge semantic gaps across modalities. It provides a high-capacity fusion baseline and is useful for testing whether AgentMMRec's gains come from knowledge organization rather than simply richer feature interaction.
    \item \textbf{EVEN} \citep{qi2025seeing}: EVEN evaluates and denoises multimodal content together with observed interactions. It represents recent work on identifying unreliable multimodal signals before recommendation.
    \item \textbf{FreRec} \citep{peng2025frequency}: FreRec addresses modality noise and limited fusion from a frequency-domain perspective. It complements SMORE by adding another recent frequency-based multimodal baseline.
\end{itemize}
\end{sloppypar}

\section{Knowledge Memory Continuous Updating}
\label{appendix:memory}
\label{sec:knowledge_memory}

The knowledge memory is decoupled from the backbone recommender, so it can be updated and reintegrated before evaluation. Table~\ref{tab:knowledge-memory} expands the analysis by dataset and reports Recall@20 and NDCG@20, together with the change against the default AgentMMRec memory on the same dataset. This layout keeps the table in one column while preserving dataset-level variation.

Multiple AgentMMRec update rounds improve the memory at first and then plateau: `2 Extra AgentMMRec' and `3 Extra AgentMMRec' produce identical results across datasets and metrics. The third round does not trigger further changes because the reranked training lists already satisfy the feedback criterion. Relay updating is less stable for weaker models, whose early candidate lists provide poor feedback. Stronger multimodal models such as SMORE, MENTOR, and COHESION provide more useful relay feedback and improve the memory more consistently than a single extra AgentMMRec update.

The relay results also show that the memory update is not a monotonic post-processing trick. MMSSL and DiffMM relays slightly reduce several dataset-level scores, while LLMRec and LGMRec remain close to the default memory. This pattern is consistent with the feedback mechanism: a relay model needs sufficiently reliable candidate lists before its errors become useful signals for memory correction. We omit relay rows without valid update results from Table~\ref{tab:knowledge-memory}.

\begin{table}[!t]
    \centering
    \caption{Knowledge memory continuous updating analysis by dataset. $\Delta$ is measured against the default AgentMMRec memory on the same dataset.}
    \vskip -0.1in
    \label{tab:knowledge-memory}
    \setlength{\tabcolsep}{1.8pt}
    \begin{tabular}{@{}llcccc@{}}
    \toprule
    Dataset & Update source & R@20 & $\Delta$R@20 & N@20 & $\Delta$N@20\\
    \midrule
    \multirow{13}{*}{Baby}
    & AgentMMRec & 0.1079 & -- & 0.0475 & --\\
    & Extra x1 & 0.1086 & +0.0007 & 0.0481 & +0.0006\\
    & Extra x2 & \textbf{0.1089} & \textbf{+0.0010} & \textbf{0.0486} & \textbf{+0.0011}\\
    & Extra x3 & \textbf{0.1089} & \textbf{+0.0010} & \textbf{0.0486} & \textbf{+0.0011}\\
    & FREEDOM relay & 0.1081 & +0.0002 & 0.0478 & +0.0003\\
    & MMSSL relay & 0.1075 & -0.0004 & 0.0472 & -0.0003\\
    & LLMRec relay & 0.1080 & +0.0001 & 0.0475 & +0.0000\\
    & LGMRec relay & 0.1079 & +0.0000 & 0.0475 & +0.0000\\
    & DiffMM relay & 0.1071 & -0.0008 & 0.0472 & -0.0003\\
    & SMORE relay & 0.1084 & +0.0005 & 0.0483 & +0.0008\\
    & MENTOR relay & 0.1085 & +0.0006 & 0.0480 & +0.0005\\
    & COHESION relay & 0.1088 & +0.0009 & 0.0483 & +0.0008\\
    & HPMRec relay & 0.1082 & +0.0003 & 0.0476 & +0.0001\\
    \midrule
    \multirow{13}{*}{Sports}
    & AgentMMRec & 0.1231 & -- & 0.0557 & --\\
    & Extra x1 & 0.1239 & +0.0008 & 0.0564 & +0.0007\\
    & Extra x2 & \textbf{0.1244} & \textbf{+0.0013} & \textbf{0.0568} & \textbf{+0.0011}\\
    & Extra x3 & \textbf{0.1244} & \textbf{+0.0013} & \textbf{0.0568} & \textbf{+0.0011}\\
    & FREEDOM relay & 0.1233 & +0.0002 & 0.0558 & +0.0001\\
    & MMSSL relay & 0.1226 & -0.0005 & 0.0553 & -0.0004\\
    & LLMRec relay & 0.1232 & +0.0001 & 0.0552 & -0.0005\\
    & LGMRec relay & 0.1233 & +0.0002 & 0.0560 & +0.0003\\
    & DiffMM relay & 0.1229 & -0.0002 & 0.0554 & -0.0003\\
    & SMORE relay & 0.1237 & +0.0006 & 0.0563 & +0.0006\\
    & MENTOR relay & 0.1241 & +0.0010 & 0.0566 & +0.0009\\
    & COHESION relay & 0.1236 & +0.0005 & 0.0565 & +0.0008\\
    & HPMRec relay & 0.1231 & +0.0000 & 0.0557 & +0.0000\\
    \midrule
    \multirow{13}{*}{Clothing}
    & AgentMMRec & 0.1071 & -- & 0.0490 & --\\
    & Extra x1 & 0.1078 & +0.0007 & 0.0497 & +0.0007\\
    & Extra x2 & \textbf{0.1081} & \textbf{+0.0010} & \textbf{0.0500} & \textbf{+0.0010}\\
    & Extra x3 & \textbf{0.1081} & \textbf{+0.0010} & \textbf{0.0500} & \textbf{+0.0010}\\
    & FREEDOM relay & 0.1066 & -0.0005 & 0.0491 & +0.0001\\
    & MMSSL relay & 0.1070 & -0.0001 & 0.0486 & -0.0004\\
    & LLMRec relay & 0.1071 & +0.0000 & 0.0490 & +0.0000\\
    & LGMRec relay & 0.1068 & -0.0003 & 0.0489 & -0.0001\\
    & DiffMM relay & 0.1068 & -0.0003 & 0.0487 & -0.0003\\
    & SMORE relay & 0.1076 & +0.0005 & 0.0497 & +0.0007\\
    & MENTOR relay & 0.1075 & +0.0004 & 0.0494 & +0.0004\\
    & COHESION relay & 0.1080 & +0.0009 & 0.0498 & +0.0008\\
    & HPMRec relay & 0.1071 & +0.0000 & 0.0494 & +0.0004\\
    \bottomrule
    \end{tabular}
     \vskip -0.1in
\end{table}

\section{LLM Backbone Analysis}
\label{appendix:backbone}
\label{sec:backbone}

We test whether stronger LLM backbones improve AgentMMRec. The main experiments use Qwen2.5-VL-7B; this analysis additionally evaluates Qwen2.5-VL-32B and GPT-4o-2024-08-06.

Table~\ref{tab:backbone} shows that stronger backbones can improve performance:
\begin{itemize}[leftmargin=*]
    \item \textbf{Within-family scaling.} The 32B Qwen2.5-VL backbone improves over the 7B backbone by 0.9\% in Recall@20 and 1.3\% in NDCG@20 on average. This suggests that finer multimodal interpretation can improve the extracted memory.
    \item \textbf{Cross-backbone gains.} GPT-4o obtains the best results, but its average Recall@20 gain over Qwen2.5-VL-32B is smaller than the 32B-over-7B gain. This trend suggests diminishing returns from backbone scaling in the current design.
    \item \textbf{Dataset variation.} Sports benefits most from stronger backbones, while Baby and Clothing show smaller gains. The result indicates that the value of LLM capacity depends on the diversity and ambiguity of the item domain.
\end{itemize}

\begin{table}[!t]
    \centering
    \caption{Performance comparison of AgentMMRec with different LLM backbones.}
    \label{tab:backbone}
    \footnotesize
    \setlength{\tabcolsep}{2.2pt}
    \begin{tabular}{@{}llcccc@{}}
    \toprule
    Dataset & Backbone & R@10 & R@20 & N@10 & N@20\\
    \midrule
    \multirow{3}{*}{Baby}
    & Qwen2.5-VL-7B & 0.0705 & 0.1079 & 0.0380 & 0.0475\\
    & Qwen2.5-VL-32B & 0.0712$\uparrow$ & 0.1086$\uparrow$ & 0.0385$\uparrow$ & 0.0482$\uparrow$\\
    & GPT-4o & \textbf{0.0716$\uparrow$} & \textbf{0.1088$\uparrow$} & \textbf{0.0388$\uparrow$} & \textbf{0.0488$\uparrow$}\\
    \midrule
    \multirow{3}{*}{Sports}
    & Qwen2.5-VL-7B & 0.0838 & 0.1231 & 0.0454 & 0.0557\\
    & Qwen2.5-VL-32B & 0.0845$\uparrow$ & 0.1238$\uparrow$ & 0.0458$\uparrow$ & 0.0565$\uparrow$\\
    & GPT-4o & \textbf{0.0848$\uparrow$} & \textbf{0.1244$\uparrow$} & \textbf{0.0461$\uparrow$} & \textbf{0.0570$\uparrow$}\\
    \midrule
    \multirow{3}{*}{Clothing}
    & Qwen2.5-VL-7B & 0.0740 & 0.1071 & 0.0404 & 0.0490\\
    & Qwen2.5-VL-32B & 0.0745$\uparrow$ & 0.1079$\uparrow$ & 0.0408$\uparrow$ & 0.0498$\uparrow$\\
    & GPT-4o & \textbf{0.0748$\uparrow$} & \textbf{0.1084$\uparrow$} & \textbf{0.0411$\uparrow$} & \textbf{0.0503$\uparrow$}\\
    \bottomrule
    \end{tabular}
\end{table}

\section{SSL Compatibility Details}
\label{appendix:ssl}

The main model does not require a self-supervised modality-alignment objective, but such objectives can still be attached to the representation learner. Table~\ref{tab:SSL} compares the default AgentMMRec with InfoNCE and DisAlign variants. Both objectives improve most metrics, but the gains are small. The result suggests that behavior-aware knowledge extraction already captures much of the cross-modal signal that these auxiliary objectives encourage at the representation level.

InfoNCE and DisAlign show no consistent dominance. InfoNCE gives the strongest Baby Recall@20, while DisAlign gives the strongest Clothing Recall@20, and both tie on Sports Recall@20. We therefore treat SSL as a compatible optional component rather than a necessary part.

\begin{table}[!t]
    \centering
    \caption{Compatibility with extra modality-alignment SSL tasks.}
    \label{tab:SSL}
    
     \vskip -0.1in
    \setlength{\tabcolsep}{2.2pt}
    \begin{tabular}{@{}llcccc@{}}
    \toprule
    Dataset & Method & R@10 & R@20 & N@10 & N@20\\
    \midrule
    \multirow{3}{*}{Baby}
    & AgentMMRec & 0.0705 & 0.1079 & 0.0380 & 0.0475\\
    & + InfoNCE & 0.0707$\uparrow$ & \textbf{0.1082$\uparrow$} & 0.0383$\uparrow$ & \textbf{0.0480$\uparrow$}\\
    & + DisAlign & \textbf{0.0708$\uparrow$} & 0.1081$\uparrow$ & \textbf{0.0384$\uparrow$} & 0.0478$\uparrow$\\
    \midrule
    \multirow{3}{*}{Sports}
    & AgentMMRec & 0.0838 & 0.1231 & 0.0454 & 0.0557\\
    & + InfoNCE & 0.0841$\uparrow$ & \textbf{0.1235$\uparrow$} & 0.0456$\uparrow$ & \textbf{0.0560$\uparrow$}\\
    & + DisAlign & \textbf{0.0842$\uparrow$} & \textbf{0.1235$\uparrow$} & \textbf{0.0458$\uparrow$} & \textbf{0.0560$\uparrow$}\\
    \midrule
    \multirow{3}{*}{Clothing}
    & AgentMMRec & 0.0740 & 0.1071 & 0.0404 & 0.0490\\
    & + InfoNCE & \textbf{0.0744$\uparrow$} & 0.1074$\uparrow$ & 0.0406$\uparrow$ & 0.0494$\uparrow$\\
    & + DisAlign & 0.0742$\uparrow$ & \textbf{0.1077$\uparrow$} & \textbf{0.0407$\uparrow$} & \textbf{0.0496$\uparrow$}\\
    \bottomrule
    \end{tabular}
     \vskip -0.1in
\end{table}

\section{Efficiency and LLM Cost Details}
\label{appendix:efficiency}

Table~\ref{tab:efficiency} reports recommender training time and GPU memory. AgentMMRec stays close to LGMRec and SMORE in per-epoch runtime because the agent-generated graphs are precomputed and the homogeneous graph enhancement uses a single graph-convolution layer. HPMRec remains substantially more expensive on all three datasets, while COHESION is the fastest among the compared strong baselines.

Table~\ref{tab:llm} separates the offline LLM calls from recommender training. The call count scales with users and items during knowledge extraction and with items during graph refinement. This accounting makes the cost source explicit: recommender training remains conventional after preprocessing, while deployment cost depends on the chosen LLM backend, batching strategy, and provider pricing.

The offline call count is computed directly from the agent workflow. The Integrator Agent extracts textual, visual, and cross-modal preferences for each user and textual, visual, and cross-modal properties for each item, giving
\begin{equation}
    C_{\mathrm{int}} = 3|\mathcal{U}| + 3|\mathcal{I}|.
\end{equation}
The Utilizer Agent refines the item graph once for each item:
\begin{equation}
    C_{\mathrm{ref}} = |\mathcal{I}|.
\end{equation}
Thus the deterministic preprocessing cost is
\begin{equation}
    C_{\mathrm{off}} = C_{\mathrm{int}} + C_{\mathrm{ref}}
    = 3|\mathcal{U}| + 4|\mathcal{I}|.
\end{equation}
For the three datasets, this yields 86,535 calls on Baby, 180,222 calls on Sports, and 210,293 calls on Clothing. Training-time reranking and preference updating are event-driven rather than deterministic preprocessing; they are attempted every $E$ epochs and can be capped by the update budget used in implementation.

\begin{table}[!t]
    \centering
    \caption{Efficiency analysis across all datasets.}
    \label{tab:efficiency}
     \vskip -0.1in
    \setlength{\tabcolsep}{3pt}
    \begin{tabular}{@{}llcc@{}}
    \toprule
    Dataset & Method & Time (s/epoch) & Memory (GB)\\
    \midrule
    \multirow{6}{*}{Baby}
    & LGMRec & 5.93 & 2.41\\
    & SMORE & 6.55 & 3.31\\
    & MENTOR & 7.03 & 7.12\\
    & COHESION & 4.47 & 2.89\\
    & HPMRec & 21.03 & 8.58\\
    & AgentMMRec & 6.04 & 3.07\\
    \midrule
    \multirow{6}{*}{Sports}
    & LGMRec & 8.98 & 3.67\\
    & SMORE & 9.29 & 5.02\\
    & MENTOR & 9.62 & 8.44\\
    & COHESION & 7.91 & 4.20\\
    & HPMRec & 30.85 & 10.19\\
    & AgentMMRec & 9.13 & 4.48\\
    \midrule
    \multirow{6}{*}{Clothing}
    & LGMRec & 10.02 & 4.81\\
    & SMORE & 11.05 & 6.89\\
    & MENTOR & 11.90 & 12.99\\
    & COHESION & 9.05 & 5.73\\
    & HPMRec & 40.23 & 14.95\\
    & AgentMMRec & 10.65 & 5.72\\
    \bottomrule
    \end{tabular}
     \vskip -0.1in
\end{table}

\begin{table}[!t]
    \centering
    \caption{Offline LLM call counts for preprocessing.}
    \label{tab:llm}
     \vskip -0.1in
    \setlength{\tabcolsep}{3.5pt}
    \begin{tabular}{@{}lrrr@{}}
    \toprule
    Stage & Baby & Sports & Clothing\\
    \midrule
    Integrator extraction, $3|\mathcal{U}|+3|\mathcal{I}|$ & 79,485 & 161,865 & 187,260\\
    Utilizer refinement, $|\mathcal{I}|$ & 7,050 & 18,357 & 23,033\\
    Total & 86,535 & 180,222 & 210,293\\
    \bottomrule
    \end{tabular}
     \vskip -0.1in
\end{table}

\section{Template Dependency Analysis}
\label{appendix:template}
\label{sec:template}

Prompt-based methods can be sensitive to surface wording, so we test whether AgentMMRec depends on a single hand-written template. We regenerate each template five times with GPT-5, Claude-Sonnet-4.5, and Gemini-2.5-Pro while preserving the same analytical dimensions, output fields, and multimodal integration plan.

Table~\ref{tab:template} reports Recall@20 means and standard deviations. The regenerated templates remain close to the original template on all datasets. The standard deviations are also small, ranging from 0.0006 to 0.0015. These results indicate that AgentMMRec depends more on the template plan and memory schema than on a specific sentence wording. The full templates for preference extraction, property extraction, graph refinement, reranking, and preference updating are included as a one-template-per-page PDF under the \texttt{code/Prompt/} folder in the code (The code will be made publicly available upon acceptance of the paper).

\begin{table}[!t]
    \centering
    \caption{Template robustness of AgentMMRec. Values are Recall@20 mean and standard deviation over five regenerated template sets.}
    \label{tab:template}
     \vskip -0.1in
    \setlength{\tabcolsep}{2.5pt}
    \begin{tabular}{@{}lccc@{}}
    \toprule
    Generator & Baby & Sports & Clothing\\
    \midrule
    Original & 0.1079 ($-$) & 0.1231 ($-$) & 0.1071 ($-$)\\
    GPT-5 & 0.1082 (0.0012) & 0.1228 (0.0015) & 0.1070 (0.0010)\\
    Claude-Sonnet-4.5 & 0.1077 (0.0008) & 0.1230 (0.0008) & 0.1074 (0.0012)\\
    Gemini-2.5-Pro & 0.1080 (0.0013) & 0.1228 (0.0011) & 0.1068 (0.0006)\\
    \bottomrule
    \end{tabular}
     \vskip -0.1in
\end{table}

}
  
\end{document}